\documentclass[manuscript]{aastex}

\shorttitle{cyclotron variations}
\shortauthors{Maitra \& Paul}

\begin{document}

\title{Pulse phase dependent variations of the cyclotron
absorption features of the accreting pulsars
A 0535+26, XTE J1946+274 and 4U 1907+09 with Suzaku}

\author{Chandreyee Maitra\altaffilmark{1} and Biswajit Paul}
\affil{Raman Research Institute, Sadashivnagar, Bangalore-560080, India }
\email{cmaitra@rri.res.in; bpaul@rri.res.in}
\altaffiltext{1}{Joint Astronomy Programme, Indian Institute of Science, Bangalore-560012, India}

\begin{abstract}
 We have performed a detailed pulse phase resolved spectral analysis of the
 cyclotron resonant scattering features (CRSF) of the
 two Be/X-ray pulsars A0535+26 and XTE J1946+274 and the wind accreting HMXB pulsar 4U 1907+09 using
\emph{Suzaku} observations. The CRSF parameters vary strongly 
over the pulse phase and can be used to map the magnetic field 
and a possible deviation form the dipole geometry in these sources. It
also reflects the conditions at the
accretion column and the local environment over the changing viewing angles.
The pattern of variation 
with pulse phase are obtained with more than one continuum spectral
models for each source, all of which give consistent results. Care is also taken to perform the analysis over a stretch of data having 
constant spectral characteristics and luminosity to ensure that the
results reflect the variations due to the changing viewing angle alone. For A0535+26 and XTE J1946+274 which show
energy dependent dips in their pulse profiles, a partial covering
absorber is added in the continuum spectral models to take into
account an additional absorption at those phases by the accretion stream/column blocking our line of sight.
\end{abstract}


\section{Introduction}

The X-ray binary sources in which the compact object is a highly magnetized neutron star, often with a massive companion are called 
accretion powered pulsars. Due to the strong magnetic field of the neutron star, the matter here 
flows along the magnetic field lines to the poles of the system, forming an X-ray
emitting accretion column above it \citep{pringle1972,davidson1973,lamb1973}. 
Another important consequence of
the strong magnetic fields ($\sim 10^{12}$ G) are the
cyclotron resonant scattering features (CRSFs) formed by the
resonant scattering of photons by the electrons which are quantized into Landau levels forming
absorption like features at
multiples of
$\ensuremath{E_{c}}=11.6 keV\times\frac{1}{1+z}\times\frac{B}{10^{12}G}$, \ensuremath{E_{c}} being the
centroid energy,
z  the gravitational redshift
and B the magnetic field strength of the neutron star. The CRSFs
thus provide a direct tool to measure the magnetic field strength of the
neutron star. It was first discovered in the spectrum of Her X-1 \citep{trumper1977,trumper1978} and about 20
sources with CRSFs have been 
discovered so far \citep{potts2012}. The CRSFs
 which are found mostly 
in high mass X-ray binaries,  about a half 
of which are transient sources, lie between the energy range of 10-60 keV. In addition to the magnetic field 
strength, the CRSFs also provide  
crucial information on the emission geometry and its physical parameters like the electron temperature, 
optical depth etc. \\
 Pulse phase resolved spectroscopy of the cyclotron parameters is an
especially useful tool to probe the emission geometry at different viewing angle as the neutron star rotates. 
It can  further be used to map
the magnetic field geometry of the neutron star. Since the CRSFs also
show variations with luminosity and spectral changes, to perform pulse
phase resolved analysis, care should be taken to obtain the results solely due to the changing 
viewing angle by averaging over the data
stretch with similar counts and spectral ratios.  Proper continuum modeling of the 
energy spectrum also plays an important role in phase resolved analysis. \emph{Suzaku}, with its 
broadband energy coverage is most ideally suited in this regard.

A0535+26 is a Be/X-ray binary pulsar which was discovered during a 
giant outburst in 1975 by \emph{Ariel V} \citep{rosenberg1975}. It consists of a 103 s pulsating neutron star 
with a O9.7IIIe optical companion HDE245770 \citep{bartolini1978} in an eccentric orbit of e=0.47, with 
orbital period of 111 days \citep{finger2006}. The distance to the source is $\sim$ 2 kpc 
\citep{giangrande1980,steele1998}. Upto six
giant outbursts have been detected in this source so far, the latest ones during 2009/2010 
\citep{caballero2011,caballero2011a}.
The last giant outburst was followed by two smaller outbursts with a periodicity
of 115 days which is longer than its orbital period. Precursors to the giant outburst was also 
observed with the same periodicity \citep{mihara2010}.
CRSFs at $\sim$ 45 keV and $\sim$ 100 keV were discovered in this source during the 1989 giant 
outburst with \emph{HEXE} \citep{kendziorra1994}.
The second harmonic at $\sim$ 110 keV was confirmed with \emph{OSSE} during the 1994 
outburst \citep{grove1995}, although the presence of the 
fundamental at $\sim$ 45 keV was dubious. It was later confirmed during the 2005 
outburst with \emph{Integral}, \emph{RXTE} 
\citep{caballero2007} and \emph{Suzaku} observations \citep{terada2006}.

XTE J1946+274 is a transient Be/X-ray binary pulsar discovered by ASM
onboard \emph{RXTE} \citep{smith1998}, and CGRO onboard \emph{BATSE}
\citep{wilson1998} during a giant outburst in 1998, revealing 15.8 s pulsations. The optical 
counterpart was identified as an 
optically faint  B $\sim$ 18.6 mag, bright infrared (H $\sim$ 12.1)
 Be star \citep{verrecchia2002}. 
The source has a moderately eccentric orbit of 0.33 with an orbital period of 169.2 days
\citep{paul2001,wilson2003}. After the initial 
giant outburst and several short outbursts at periodic intervals, the source went into quiescence for a long
time until the recent outburst in 2010 \citep{caballero2010}. A
CRSF was discovered at $\sim$ 35 keV from the \emph{RXTE} data of the
1998 outburst observations \citep{heindl2001}.

4U 1907+09 is a persistent wind accreting high mass X-ray binary discovered in the 
\emph{Uhuru} surveys \citep{giacconi1971,schwartz1972}, having a highly redenned companion star 
(O8-O9 Ia) of magnitude 16.37 mag and a mass loss rate of $\dot{M}=7*10^{-6} M\odot yr^{-1}$ 
\citep{cox2005}. It has a moderately eccentric (e=0.28) orbit of 8.3753 days \citep{intzand1998}. It is a slowly
rotating neutron star with period of $\sim$ 440 s, and has showed several episodes of torque reversals 
with a steady spin-down from 1983 to 1998 \citep{intzand1998,mukherjee2001}, a much slower spin-down from 1998 to 2003
\citep{baykal2006}, a torque reversal between 2004 and 2005 \citep{fritz2006} and a second torque reversal 
between 2007 and 2008 \citep{inam2009}, which restored the source to the same spin-down rate before 1998.
A CRSF at $\sim$ 19 keV was reported using data from the \emph{Ginga} /observations
\citep{makishima1992, makishima1999} and was later confirmed from the \emph{BeppoSaX} observations 
with the discovery of a harmonic at $\sim$ 36 keV. \citep{cusumano1998}. \cite{rivers2010} performed a time and phase
resolved analysis of 
the \emph{Suzaku} observations of the source made during 2006 and 2007. \\
Here we present the results obtained from a pulse phase resolved spectroscopic analysis of these three 
sources with a motivation to investigate the variation pattern of the cyclotron parameters with pulse phase. 
The pulse phase dependence
of the CRSF parameters are presented for the first time for A0535+26 and XTE J1946+274, whereas a
more detailed result is presented for 4U 1907+09 which is in agreement
with the earlier results of \cite{rivers2010}. The analysis is done 
 taking into account various factors which might smear the pulse phase dependence results.The results 
 presented in this paper are one of the most detailed results
on pulse phase resolved measurements of CRSF available so far. 

\section{Observations \& Data Reduction}

There are two sets of scientific instruments onboard Suzaku. The X-ray
Imaging Spectrometer XIS \citep{koyama2007} consisting of three front
illuminated CCD detectors (FI :XIS0, XIS2, XIS3) and one back illuminated
CCD detector (BI: XIS1) work in the 0.2-12 keV range and the Hard X-ray
Detector (HXD) made of PIN diodes \citep{takahashi2007} and GSO crystal
scintillator detectors cover the energy bands of 10-70 keV and 70-600 keV
respectively. \\

\emph{Suzaku} \citep{mitsuda2007} observed A0535+26 twice, once on
September 14--15 2005 during the decline of the second normal outburst of 2005, and again on August 24 2009 
during the decline of the 2009 normal outburst. We have chosen the 2009 observation (Obs. Id--404054010) 
for our analysis because of the longer duration (exposure $\sim$ 52 ks) and its 'HXD nominal' pointing 
position which is more suitable for CRSF studies, although the count rates were comparable for both the 
observations. The XIS's were operated in the '$\frac{1}{4}$ window' 'burst' clock data mode 
which has a total time resolution of 2 s. \\
\quad XTE J1946+274 was observed on October 11 2010 (Obs. Id--405041010)
just after the peak of the September/October 2010 normal outburst.
The source was observed for  $\sim$ 51 ks in the 'HXD nominal' pointing position, and the XIS's were 
operated in the '$\frac{1}{4}$ window' 'normal' clock data mode 
which has a time resolution 
of 2 s.\\
4U 1907+09 was also observed twice with \emph{Suzaku}, once on May 2006, 
and again on April 2007. We have chosen the 2007 observation (Obs. 
Id--402067010) for our analysis because of similar reasons as in the 
case of A0535+26, i.e.  longer exposure of $\sim$ 158 ks and the 'HXD 
nominal' pointing position. The XIS's were operated in 'normal' clock 
data mode with no window option which has a time resolution of 8 s. 
The XIS data was reduced and extracted from the unfiltered XIS 
events, which were reprocessed with the CALDB version 20120428.
We checked for any significant photon pile-up effect in the reprocessed
XIS event files. To perform pile-up estimation, we examined the Point
Spread Function (PSF) of the XIS’s and obtained the count rate at the image peak per CCD exposure 
as given by Yamada \& Takahashi \citep{yamada2012}\footnote{http://www-utheal.phys.s.u-
tokyo.ac.jp/~yuasa/wiki/index.php\\/How\_to\_check\_pile\_up\_of
\_Suzaku\_XIS\_data}. Crab data is assumed to be free from pile-up
and has a value of 36 \ensuremath{ct/ sq\,  arcmin /s/CCD\, exposure} at the image peak.
Following their procedure, the XIS data of A0535
+26 and XTE J1946+274 had values of 2-3 \ensuremath{ct/ sq\, arcmin /s/CCD\, exposure} at the image peaks,
and showed no evidence of significant pile-up.
4U 1907+09 on the other hand showed a case of moderate photon
pile-up. The value obtained at the image center was higher than the 
 Crab Nebula count rate of 36 \ensuremath{ct/ sq\,  arcmin /s/CCD\, exposure}.  The 
radius at which this value equals 36 in the PSF is about 15-16 arcsec, and hence 15 pixels were removed
from the image center to account 
for this effect.
For the extraction of XIS light curves and spectra from the
reprocessed XIS data, a $4^{'}$ diameter circular region
was selected around the source centroid for A0535+26 and XTE J1946
+274, and an additional central 15 ($\sim$ $16^{''}$) 
pixels were removed in the case of 4U 1907+09 to discard the maximum pile up affected regions.
Background light curves and spectra were also extracted by selecting  
 regions of the same size away from the source.
The XIS count rate was 3.6 c/s, 3.1 c/s and 8.1 c/s for A0535+26, XTE J1946
+274 and 4U 1907+09 respectively. 4U 1907+09 had a $\sim 12 $ \% loss in count rate after the removal of the photons from the 
central region due to the
pile-up correction. Response files and effective area files were 
generated by using the 
FTOOLS task 'xisresp'. The HXD/PIN light curves and 
spectra were extracted after reprocessing  the  unfiltered event files
\footnote{http://heasarc.nasa.gov/docs/suzaku/analysis/hxd
\_repro.html}.
The HXD/PIN background was created by adding the simulated 'tuned' non X-ray background event 
files 
(NXB) corresponding to the month and year of the respective observations 
\cite{fukazawa2009}\footnote{http://heasarc.nasa.gov/docs/suzaku/analysis/
pinbgd.html} to the the cosmic X-ray 
background, which was simulated as suggested by the instrument team\footnote
{http://heasarc.nasa.gov/docs/suzaku/analysis/pin\_cxb.html} after applying 
appropriate 
normalizations for both cases. The corresponding response files were obtained from 
the \emph{Suzaku} guest observatory facility.\footnote{http://
heasarc.nasa.gov/docs/heasarc/caldb/suzaku/ and used for the HXD/PIN 
spectrum}  

\section{A 0535+26 and XTE J1946+274}
\subsection{Timing analysis: Light curves \& Hardness ratio \& pulse
period determination}
We performed timing analysis after applying barycentric corrections to the 
event data files using the FTOOLS 
task 'aebarycen'. Light curves were extracted with a time resolution of 2 s for the XISs (0.2--12 keV), and 1 s 
for the HXD/PIN (10--70 keV) respectively. For XTE J1946+274 which has a short
pulse period, light curve with the resolution of 10 ms was extracted from the  
HXD/PIN data to search for the pulse period. We applied pulse folding and $\chi^{2}$ maximization 
technique to search for
pulsations in the XIS data for A 0535+26 and PIN/HXD data for XTE J1946+274.
The best estimate of the period was found to be $103.47 \pm 0.09$ s 
for A0535+26. This value is consistent with the pulse period 
determined from the \emph{INTEGRAL IBIS} data during the same outburst 
at MJD 55054.995 \citep{caballero2010a} assuming the spin down value 
determined from the same. For XTE J1946+274, the best-fit period was
estimated to be $15.75 \pm 0.11$ s. Orbital correction of the pulse arrival times 
was not required for both the sources having a very long
orbital period.
The XIS and PIN light curves of the sources binned with 
its pulse period for A 0535+26 and 10 pulsar periods for XTE J1946+274,
are shown in Figure~\ref{fig1}. 
The light curves show more or less 
constant count rate, and  do not have any particular trend of 
variation. For each figure, the third panel shows the hardness ratios (ratio 
of PIN counts to XIS counts) which is also more or less constant 
throughout the observation duration and does not have any signatures
of spectral variability which might affect the results of pulse phase
resolved spectroscopy. 
\subsection{Energy Dependence of the pulse profiles}
We created the energy resolved pulse profiles for the entire stretch of observations
 by folding the light curves in different energy bands 
with the obtained pulse period. The pulse profiles in the energy range of 0.3-12 keV were created 
using all the three XISs (0, 1 \& 3), and in the 10-70 keV range were created from the 
PIN data.
The energy dependence of the pulse profiles in A 0535+26 are shown in  Figure~\ref
{fig3}. The pulse profiles are complex in structure with narrow dips in the low energy ranges
$\le 12$ keV which morphed to become a simpler, more sinusoidal profile at higher energies.
The following characteristics are
observed with a careful examination of the profiles.

\begin{enumerate}
\item A narrow dip at phase $\sim$ 0.1, which decreases in strength with energy and disappears at
energies $\ge 14$ keV. 
\item Indication of another sharp dip at phases  $\sim$ 0.2--0.3,
which is evident only at the lowest energy range ($\le 2 $keV).
 \item The emission component between phases 0.5--0.7 becomes weaker
and weaker with energy and finally disappears at $\sim$ 17 keV. As a result the main dip of the 
profile (at phase $\sim 0.6$) is narrower at lower
energies ($\le 12$ keV) and broader at higher energies
\end{enumerate}
The energy dependence is very similar to that found during the 2005 \emph{Suzaku} observation \citep{naik2008}. The profile is however, very different
from the simple sinusoidal profile at all energies found during the  quiescence phase of the source \citep{mukherjee2005,negeeruela2000},
or the double peaked profile extending upto higher energies during its giant outbursts \citep{mihara1995,kretschmar1996}.

The energy dependence of the pulse profiles of XTE J1946+274 is shown in Figure~\ref{fig4}. The pulse profiles show a clear
double peaked structure which extends upto the high energies. The following characteristics can be observed in more detail.
\begin{enumerate}
\item At the lowest energy ranges (0.3--4 keV), the peak (phase $\sim$ 0.5) increases in strength with  energy and the
dip at phase $\sim$ 0.8 increases in strength.
\item Between 4--7 keV, the same dip mentioned above decreases in strength and the two peaks are almost equal in strength.
\item Between 7--17 keV, this dip (phase $\sim$ 0.8) disappears and a new, much weaker dip appears at $\sim$ 0.9 which is probably
the true interpulse region between the pulses. 
\item At the highest energies (25--70 keV), the second peak at phase $\sim$ 0.1 becomes much weaker.
\end{enumerate}
The energy dependence of the pulse profiles of XTE J1946+274 is very similar to that investigated by \cite{wilson2003} during the 1998 outburst of the source.

\subsection{Spectroscopy}
\subsubsection{Pulse phase averaged spectroscopy}
 We performed pulse phase averaged spectral analysis 
of A 0535+26 and XTE J1946+274 using spectra from the three front illuminated CCDs
(XISs-0 and 3), the back illuminated CCD (XIS-1) and the
 PIN.
We performed spectral fitting  using \emph{XSPEC} v12.7.0. The XIS spectra were fitted from 0.8-10 keV
and the PIN spectrum from 10-70 keV. The  energy range of 1.75-2.23 keV
was neglected due to an artificial structure in the XIS spectra around 
the Si edge and Au edge. After appropriate background subtraction, the spectra were fitted 
simultaneously with all parameters tied, except the relative 
instrument normalizations 
which were kept free. The XIS spectra were rebinned by a factor of 6
from 0.8-6 keV and 7-10 keV, and by a factor of 2 between 6-7 keV. 
The PIN spectrum of A 0535+26 was rebinned by a factor of 2 upto 22 keV, by 4
upto 45 keV, and 6 upto 70 keV. Due to comparatively inferior statistics in the PIN spectrum of
XTE J1946+274, higher rebinning factors of 2, 6, and 10 were applied in the above
mentioned energy ranges.\\ 
In HMXB accretion powered pulsars, the continuum emission can be interpreted to arise by Comptonization
of soft X-rays in the plasma above the neutron
star surface. It is usually modeled phenomenologically with a powerlaw and cutoff at high energies \citep{white1983,mihara1995,coburn2001}.
The most widely used empirical models are the high energy cutoff (highecut) or the Fermi Dirac cutoff (fdcut)
\citep{tanaka1986} with the powerlaw component, or cutoff powerlaw (cutoffpl) model.
Other models include the negative-positive exponential powerlaw component (NPEX) \citep{mihara1995},
and a more physical comptonization model 'CompTT' \citep{titarchuk1994}. We tried to fit the energy spectra
with all the continuum models mentioned above, available as a standard or local 
package in \emph{XSPEC} and carried out further analysis with only the models which
gave best fits for the respective sources. \\

  A 0535+26 :
  
For A0535+26 the best fits were obtained with the NPEX, powerlaw and the 'CompTT' model (assuming spherical geometry for the comptonizing region). The powerlaw model however
did not require a 'highecut' to fit the energy spectra. Including the GSO spectra in the fitting, the relative normalization of the GSO
with respect to XIS showed that the flux in the GSO band (50-200 keV) was overestimated $\sim 4$ times without the inclusion
of a 'highecut' in the spectrum. As inclusion of the GSO spectrum is not possible for phase resolved studies due to its limited statistics, and
 a spectrum of an accretion powered pulsar without a cutoff at higher energies
is not viable, we have carried out further analysis with the 'NPEX' and 'CompTT' models. 
We applied a partial covering absorption model 'pcfabs' in both the cases along with the 
Galactic line of sight absorption, to take into account the intrinsic absorption evident at certain
pulse phases. This is evident in the pulse profiles and is a feature local to the neutron star.
The narrow Fe k$\alpha$ feature found at 6.4 keV was modeled by a gaussian 
 line. In addition, a deep and
wide feature found at $\sim$ 45 keV was modeled with a Lorentzian profile, which is the CRSF found previously in this source \citep{caballero2007}. 
For A 0535+26, the CRSF has been reported before at the same energy, even in a \emph{Suzaku} observation \citep{terada2006}.
 So we do not comment on its detection significance here.
We also tried a gaussian profile to model the CRSF feature. Since the centroid
energy of the Lorentzian description is not coincident with the minimum of the line profile \citep{nakajima2010},
apart from a slight offset between the centroid energies of the Lorentzian and gaussian profile,the other parameters 
like the depth and width are consistent between the two models. The fits are also similar. We however 
considered a Lorentzian profile for the CRSFs for the rest of the paper after verifying the consistency
between the Lorentzian and Gaussian profiles.
The CRSF parameters were also consistant within error bars
for both the continuum models, the centroid energy being only slightly higher for the 'powerlaw' model. The reduced $\chi^{2}$ obtained for the models
 were 1.25 and 1.26 for 839 and 840 d.o.f respectively with no systematic residual pattern. \\
 
 XTE J1946+274:

For XTE J1946+274, best fits with similar values of reduced $\chi^{2}$ were obtained with the 'highecut', 'NPEX' 
and 'CompTT' model. Similar to A 0535+26, the local absorption of the neutron star was taken into account
by the model 'pcfabs', and a gaussian line was also used to account for the narrow Fe k$\alpha$ feature
found at 6.4 keV. A deep and wide residual was found at $\sim$ 38 keV, at the same energy as the CRSF 
discovered by \cite{heindl2001}. As discussed previously, the CRSF was modeled with a Lorentzian profile.
The 'highecut' and 'NPEX' models gave consistant values of the CRSF parameters, but the 'CompTT' model required a much shallower and narrow profile. 
Moreover, we were unable to constrain all the parameters of the 'CompTT' well for this source, 
probably due to the poorer quality of the PIN data.
We have thus carried out the further analysis of this source with the two former models. For the best fitting models, the reduced $\chi^{2}$
was 1.09 and 1.11 respectively for 826 d.o.f. Without the inclusion of the CRSF, the difference in $\chi^{2}$ was 
150 and 119 respectively for the same models. The best-fitting values for the spectral models for both the sources are given in Table 1. 
 Figure \ref{fig5} shows the best-fit spectra for both the sources along with the
 residuals before and after including the CRSF, thus showing the presence of the feature clearly. \\

\cite{muller2012} however have reported the analysis of the \emph{RXTE,INTEGRAL} and \emph{Swift} observations 
during the same outburst of this source. Instead of a line at 36 keV, they found a weak evidence of a CRSF at $\sim$
25 keV. It may be worthwhile mentioning in this context that the \emph{Suzaku} PIN data has better sensitivity than \emph{INTEGRAL} ISGRI 
at this energy range, and hence may be better suited for CRSF detection. However we have carefully
checked the statistical significance and possible systematic errors associated the CRSF.

\emph{Statistical significance}:
To estimate the detection significance of the CRSF we tried to fit the PIN spectrum alone
with the 'highecut' model with its powerlaw index frozen to the value obtained from the best fitting broadband spectrum.
The addition of the CRSF improved the $\chi^{2}$ from 51.56 to 28.13 for 20 d.o.f corresponding 
to an F value of 16.7, and a F-test false alarm probability of $6 \times
10^{-4}$. 

\emph{Possible systematic errors}:
At first, we used the the earth occultation
data to check the reproducibility of the NXB \citep{fukazawa2009}. We extracted the spectra using the earth occultation data in three energy bands centering
the CRSF and compared ratio of the count rates with the NXB. The ratio obtained 
were 1.3, 1.2 and 1.2 at $10-28$, $28-48$ and $48-70$ keV respectively indicating the lack of any energy dependent feature that can be
introduced by the simulated X-ray background.
We also included a systematic uncertainty of $3 \%$ on the PIN spectrum to check the detection
of the CRSF. The line was still detected, but the uncertainty in the depth
of the feature increased by $23 \%$. The detection of pulse phase dependence of this feature as discussed in section 
3.3.2 is also in favor of its presence since the background data is not expected to vary over
the pulse phase. Finally, to verify the existence of the CRSF in a model independent manner, we
divided the PIN spectrum of a pulse phase with the deepest CRSF, by the same of
a pulse phase with the
shallowest CRSF detected (see section 3.3.2, pulse phase resolved
spectroscopy for the corresponding spectra). Figure \ref{fig10} shows the ratio plot of the
two spectra. Although the quality of the data is not good after 40 keV, the dip at $\sim$ 30-35 keV is clearly seen
indicating the presence of the CRSF.

\subsubsection{Pulse phase resolved spectroscopy}
For the phase-resolved analysis we extracted the source spectra for both the XIS's and the PIN data
after applying phase filtering in the
FTOOLS task XSELECT. The same background spectra and response matrices as used for the phase-
averaged spectra were however used in both the cases. 
The spectra were also fitted in the same energy range and rebinned by the same factor
as in phase-averaged case. The Galactic absorption (\ensuremath{N_{\mathrm{H1}}}) column density
and the Fe line width were frozen to the 
phase-averaged values for the two respective models. \\ 
\subparagraph{Phase resolved spectroscopy of the cyclotron parameters :} 
For investigating the pulse phase-resolved spectroscopy of the two CRSFs, 
phase resolved spectra were generated with their phases centered 
around 25 independent bins but at thrice their widths. This resulted in 25 overlapping bins out of which
only 8 were independent. We however froze the width of the CRSF to the phase-averaged value of the respective models,
and varied the rest of the continuum as well as the line parameters with pulse phase. This was due to our
inability to constrain all the parameters because of limited statistics.
Figure \ref{fig6} shows the variation of the cyclotron parameters of the sources using the 
best fit models as a function of pulse phase. 
For both the sources, the different continuum models used result in a very similar pattern 
of variation of the parameters.
This gives us a reasonable amount of confidence on the obtained results.
The following features are evident from the Figure \ref{fig6}. The results are compared with respect to the high energy
PIN profile (10--70 keV).\\
                            A 0535+26 :
\begin{enumerate}
 \item the energy (\ensuremath{E1_{\mathrm{cycl}}}) varies by $14 \%$ ($\sim$ 43--50 keV). The pattern of variation of both the energy (\ensuremath{E1_{\mathrm{cycl}}}) and depth (\ensuremath{D1_{\mathrm{cycl}}}) has a gradually increasing trend 
with the pulse profile and drops off abruptly in the off-pulse region (phase $\sim 0.6$), picking up again where the pulse profile picks up.
\item The depth (\ensuremath{E1_{\mathrm{cycl}}}) cannot be constrained at all phases by both the models, and at the off pulse phase at $\sim$ 0.6, only the 'CompTT' is able to constrain the depth.
It has a very sharp pattern of variation, varying between $\sim$ 0.8--4, and it is shallowest near the pulse peak and deepest near the pulse minima.
\end{enumerate}
                          XTE J1946+274
\begin{enumerate}

\item The energy (\ensuremath{E1_{\mathrm{cycl}}}) varies about $36 \%$. It's value is generally higher
in the first pulse with the values peaking near the first peak (phase$\sim$ 0.7--0.8), and a decreasing trend 
near the second pulse.
\item The depth (\ensuremath{E1_{\mathrm{cycl}}}) varies between 1--3. It is deepest at the interpulse regions
at phase $\sim$ 1.0 and shallow between phase 0.5--0.8 near the first peak. Due to
limited statistics, specially of the PIN spectra, the CRSF parameters however cannot be constrained at the 
main dip, and at the ascending edges of the first peak
(phase $\sim$ 0.5--0.7).
\end{enumerate}
\subparagraph{Phase resolved spectroscopy of the continuum parameters:}

A dependence of the continuum energy spectrum on the pulse phase is implied from the 
strong energy dependence of the pulse profiles, as seen in Figure~\ref{fig3} and Figure~\ref{fig4}. A 
partial covering absorption model in which the absorber is phase locked with the neutron star is required to
explain the narrow energy dependent dips in the pulse profiles. This was also our main motivation in applying the partial covering absorption 'pcfabs' to model
the continuum energy spectra. 
We generated the phase resolved spectra with 25 independent phase
bins to investigate the pulse phase-resolved spectroscopy of the continuum parameters 
 for A 0535+26. Due to the short spin period of XTE J1946+274, 25 independent phase bin extraction was not possible, specially for the XIS data.
We proceeded with the extracting of 25 overlapping but 8 independent phase bins for extraction of both XIS and PIN data as was done for the phase
resolved spectroscopy of the CRSF parameters. 
The cyclotron parameters of the corresponding phase bins were frozen to the best-fit values obtained 
from the results of investigation of the cyclotron
line parameters using 25 overlapping phase bins. Figures \ref{fig7} \&  \ref{fig8} shows phase resolved continuum parameters using the best-fit spectral models as
a function of the pulse phase for A 0535+26 and XTE J1946+274 respectively. The results obtained as seen from the Figure from both the models are as follows:
\begin{enumerate}
 \item For both the sources, there is an abrupt increase in the value of 
the local absorption component ($\ensuremath{N_{\mathrm{H2}}}$), with a corresponding change in the value of 
the  covering fraction (\ensuremath{Cv_{\mathrm{fract}}}) at the dips of the low energy XIS profile.
This picture is in agreement with a narrow stream of matter present at those phases responsible for absorption of the low
energy photons. The properties of the plasma in the accretion stream, which may be a narrow
structure having different values of opacities and optical depths can be traced from the changes in the value 
of $\ensuremath{N_{\mathrm{H2}}}$ and the covering fraction. As can also be seen clearly, the main strength of our results lie in the fact that we have obtained similar patterns
of variation of $\ensuremath{N_{\mathrm{H2}}}$ and  \ensuremath{Cv_{\mathrm{fract}}} using different continuum spectral models for the sources.
\item There are also corresponding changes in the other continuum parameters like the powerlaw photon index ($\Gamma$) of the 'powerlaw',
seed temperature (\ensuremath{CompTT_{\mathrm{T0}}}), optical depth ($\tau$) and KT of the 'CompTT' model for A 0535+26, and the powerlaw photon 
index ($\Gamma$), the E-folding and E-cut energy of the
'highecut' and the NPEX $\alpha1$ and 'KT' of the 'NPEX' model for XTE J1946+274.
The main aim of this paper is however the pulse phase resolved variation of the CRSF parameters, and detailed 
discussion of these
results are beyond the scope of this paper. 
\end{enumerate}

\section{4U 1907+09}
\subsection{Timing analysis: Light curves \& Hardness ratio \& pulse
period determination}
\begin{figure}
\begin{center}$
\begin{array}{c }
\includegraphics[scale=0.3,angle=-90]{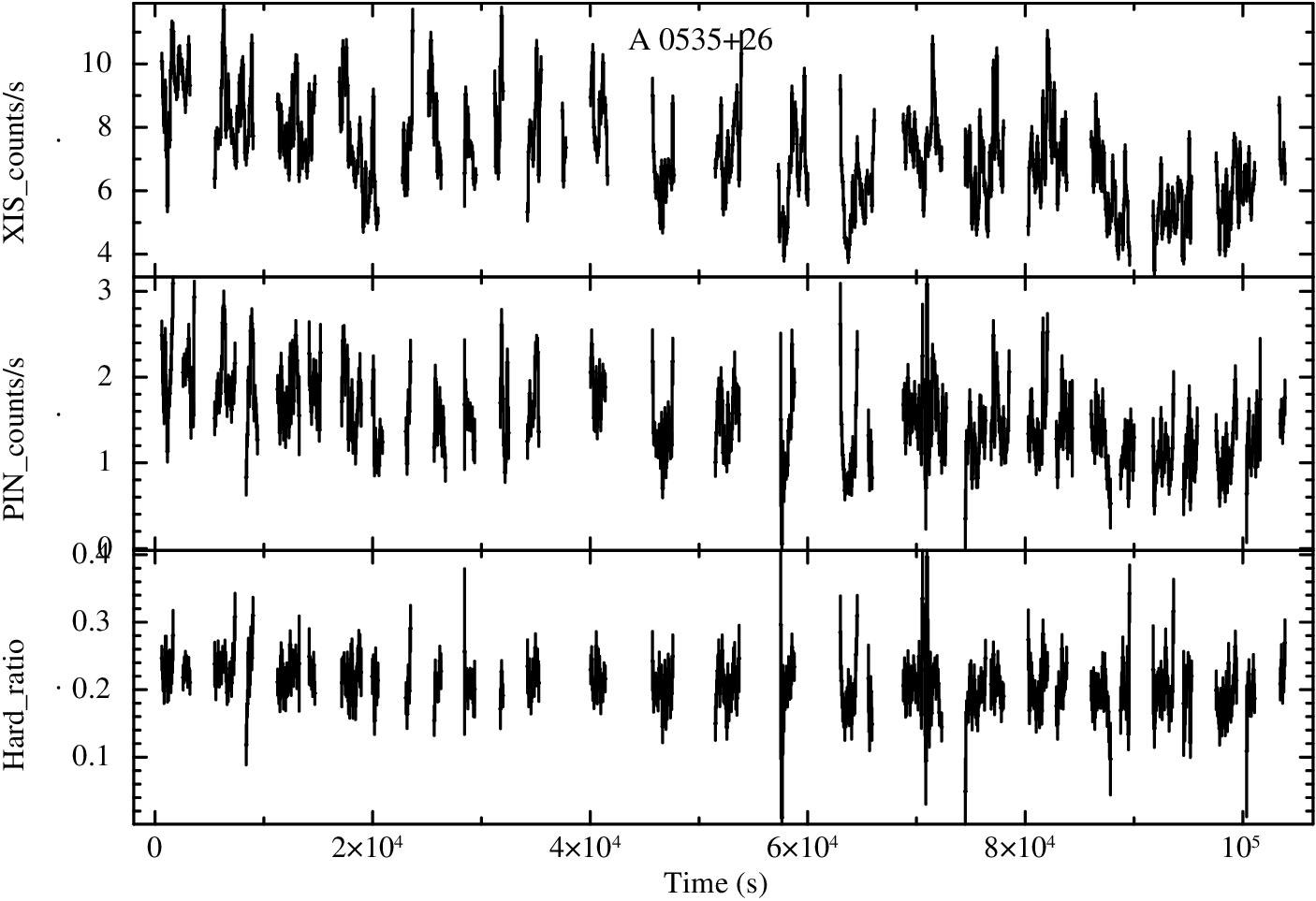} \\
\includegraphics[scale=0.3,angle=-90]{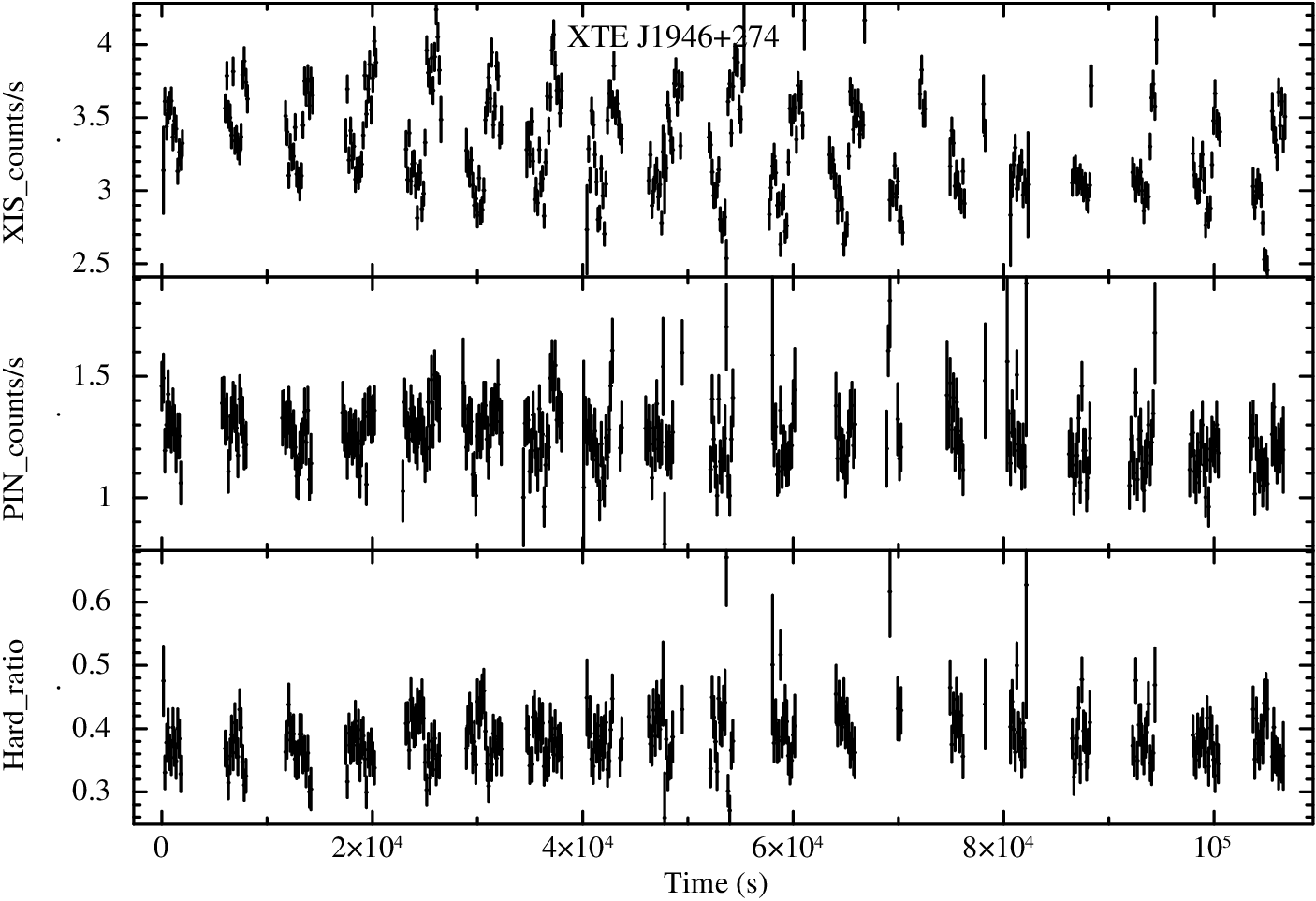} \\
\includegraphics[scale=0.3,angle=-90]{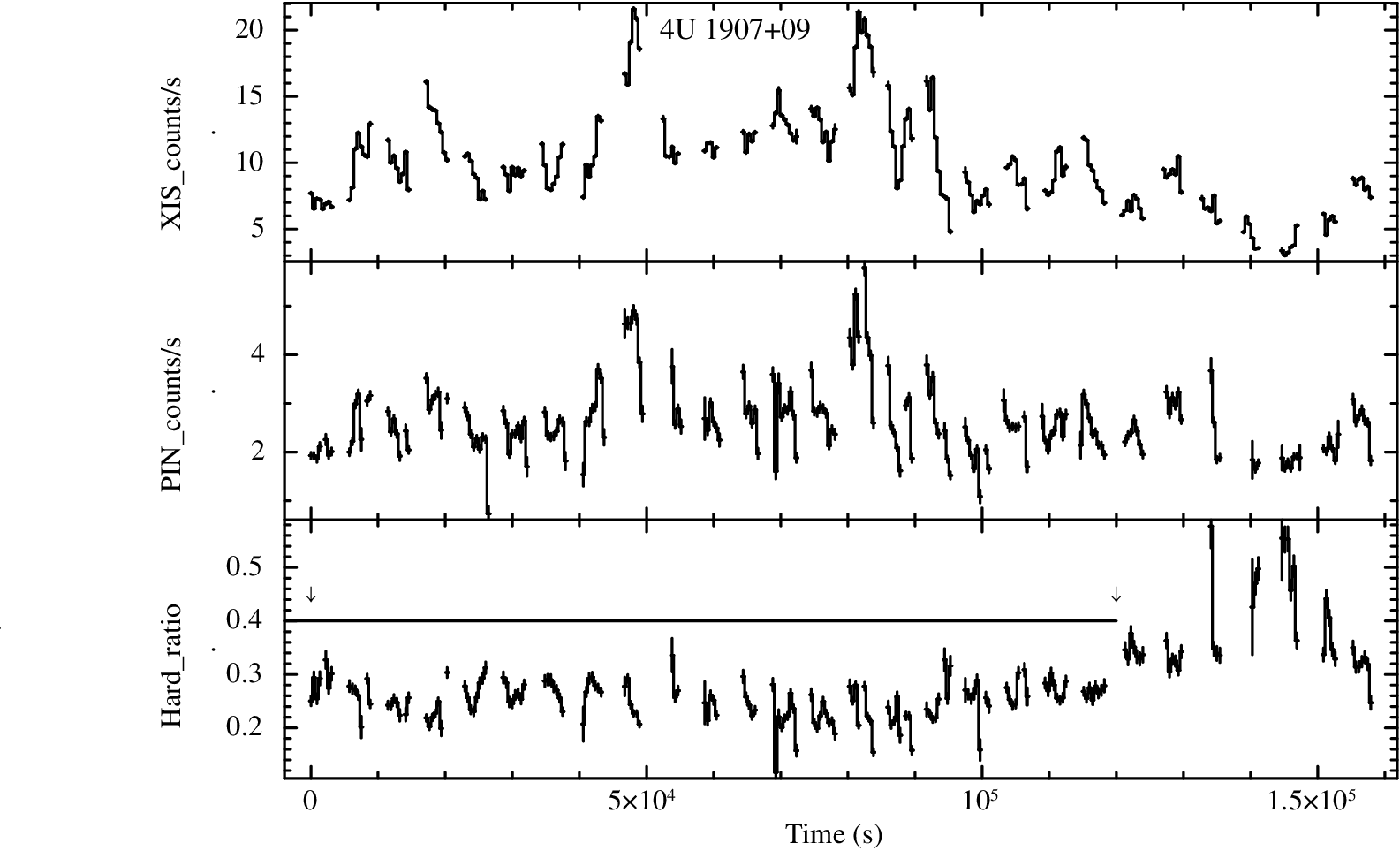} \\
\end{array}$
\end{center}
\caption{Light curves of A 0535+26, XTE J1946+274 and 4U 1907+09 obtained with
\emph{Suzaku}.The first panels in each figure shows the light curve for one of the XIS in the energy band 
of 0.3-12 keV. The second panel shows the same obtained
in the PIN energy band(10-70 keV). The time binning is equal to the respective pulse periods for A0535+26 and 4U 1907+09 and 10 pulsar
period in the case of XTE J1946+274. The bottom panel 
shows the hardness ratio. The arrows in the hardness ratio of 4U 1907+09 indicate the stretch for which data was
chosen to perform phase resolved analysis.}
\label{fig1}
\end{figure}

\begin{figure}
\begin{center}$
\begin{array}{c c}
\includegraphics[height=8 cm, width=6cm]{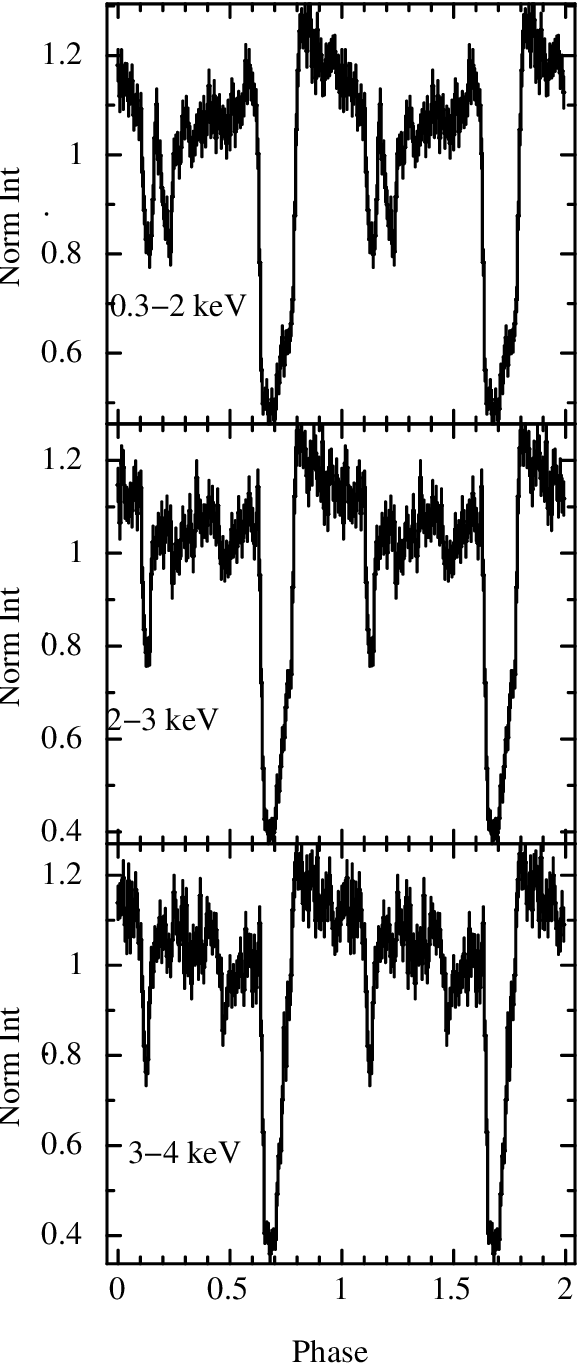} &
\includegraphics[height=8 cm, width=6cm]{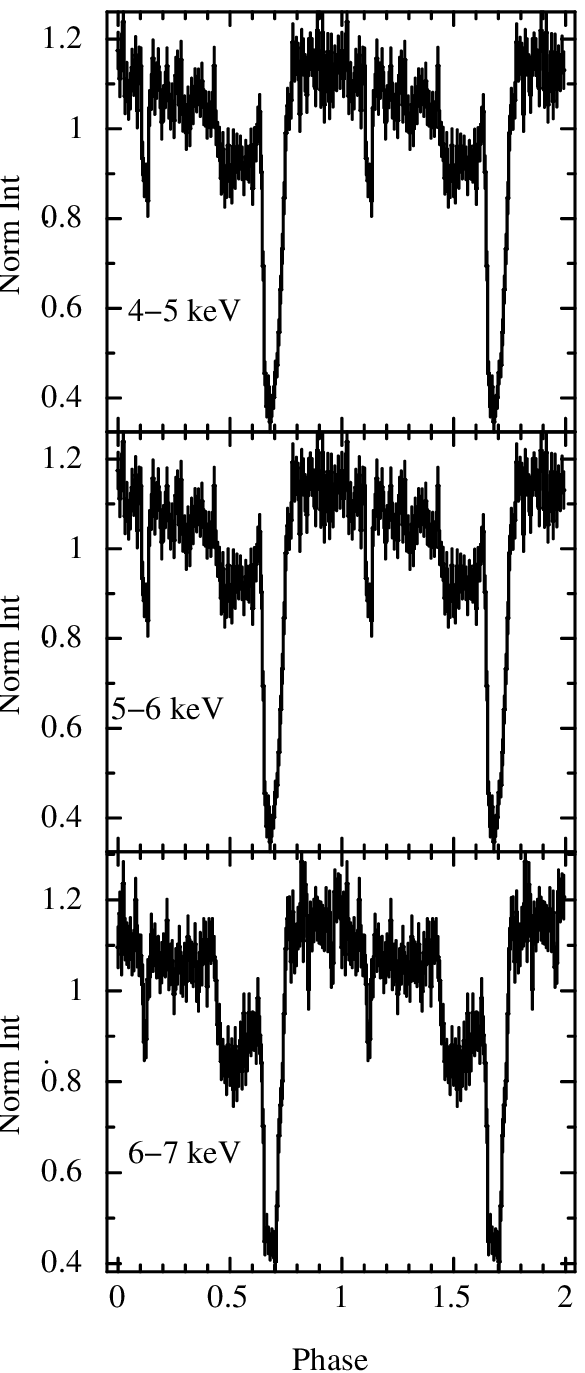}\\
\includegraphics[height=8 cm, width=6cm]{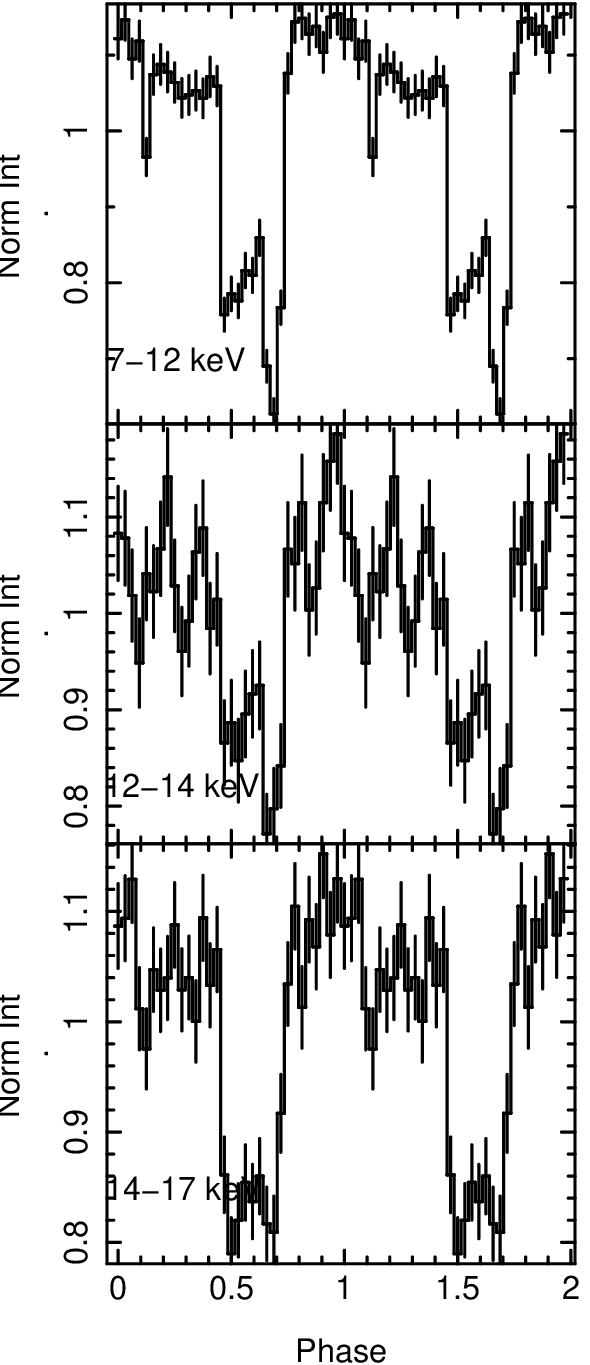} &
\includegraphics[height=8 cm, width=6cm]{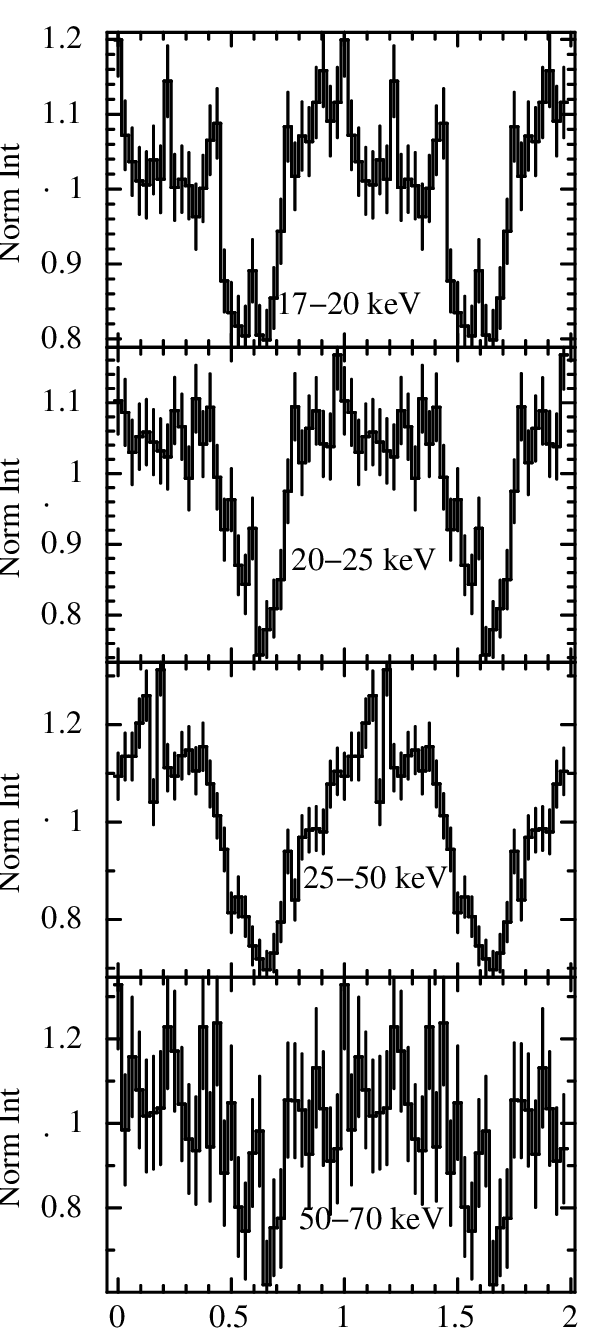} 
\end{array}$
\end{center}
\caption{Energy dependent pulse profiles of A 0535+26 using XIS \& PIN data. 
The energy range for the pulse profiles are specified inside the panels.}
\label{fig3}
\end{figure}

\begin{figure}
\begin{center}$
\begin{array}{cc}
\includegraphics[height=8 cm]{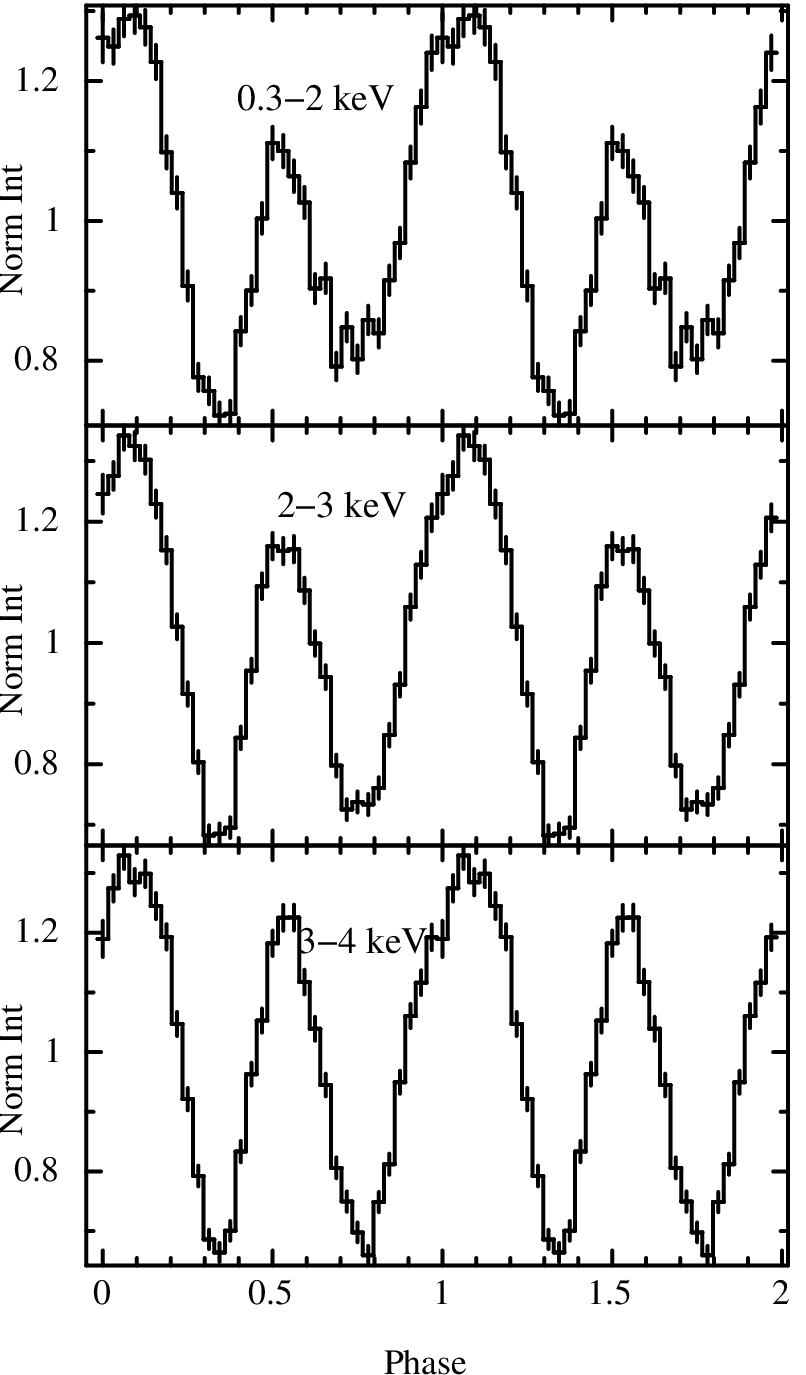} &
\includegraphics[height=8 cm]{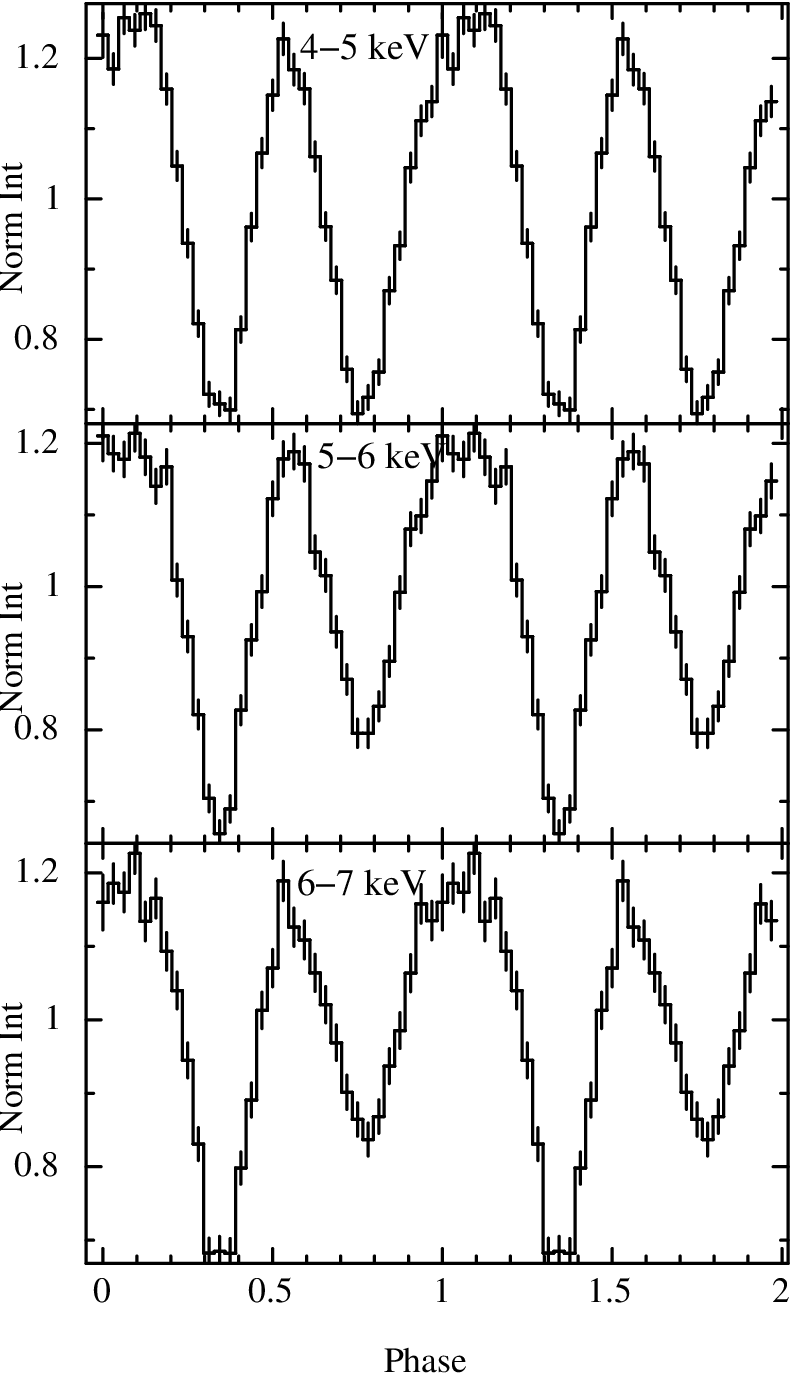}\\
\includegraphics[height=8 cm]{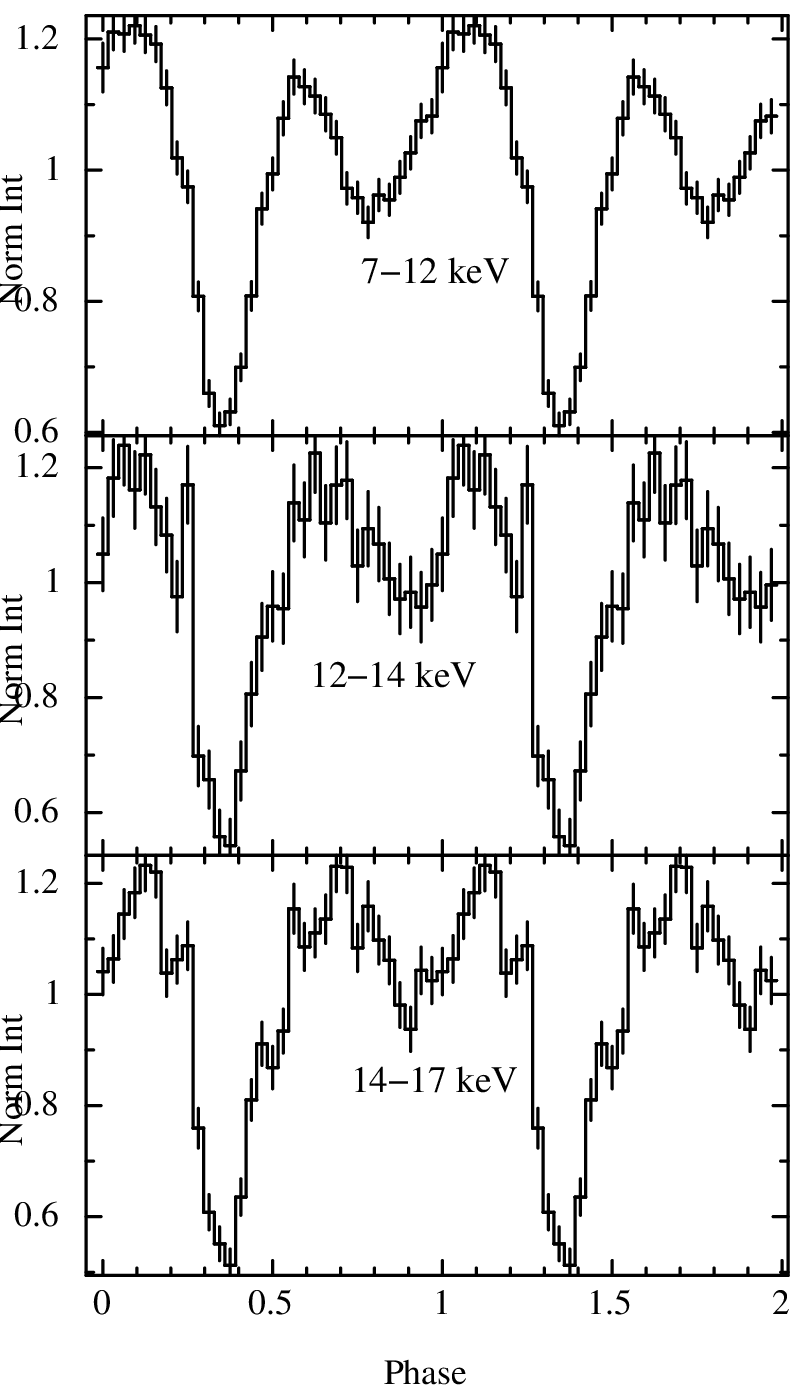} &
\includegraphics[height=8 cm]{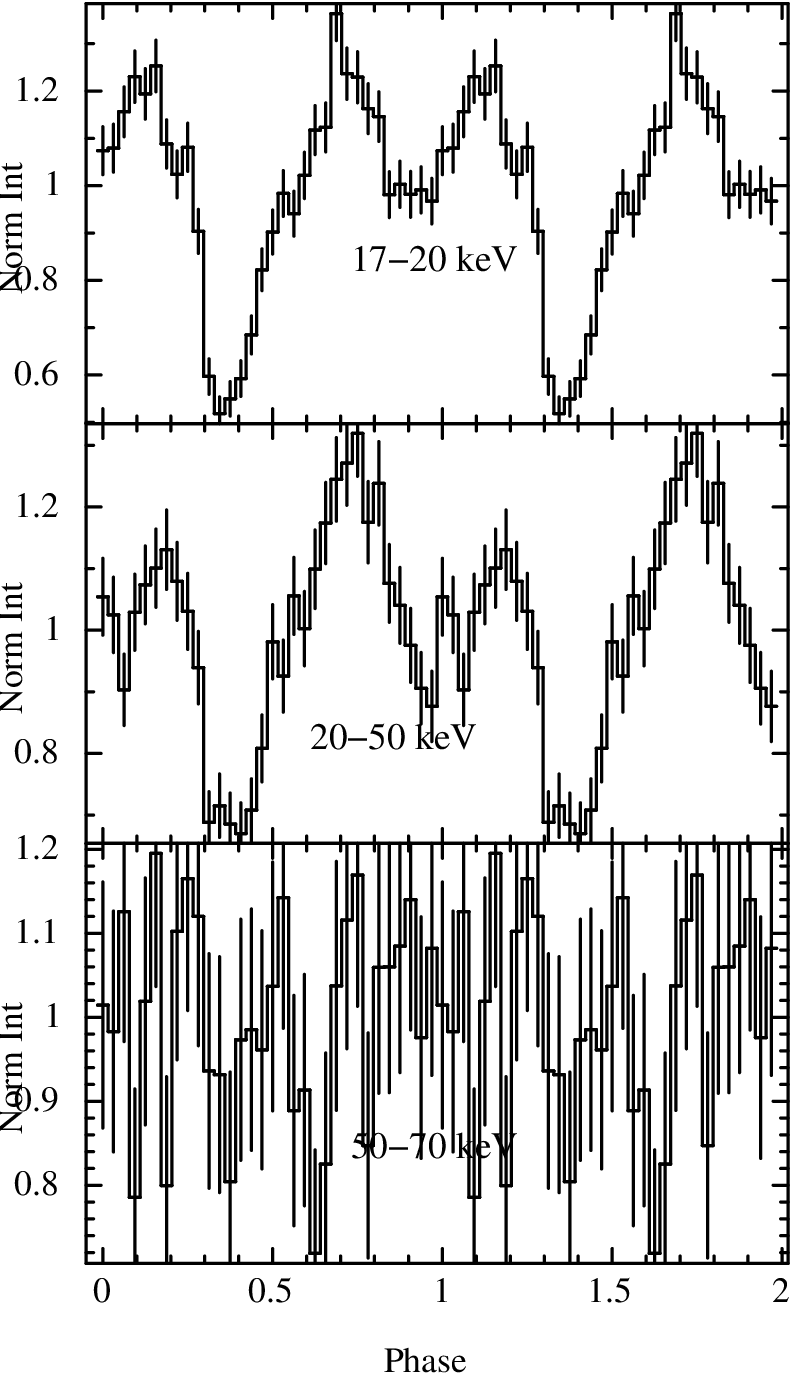} \\
\end{array}$
\end{center}
\caption{Energy dependent pulse profiles of XTE J1946+274 using XIS \& PIN data. 
The energy range for the pulse profiles are specified inside the panels.}
\label{fig4}
\end{figure}

\begin{figure}
\centering
\includegraphics[height=9cm, width=7cm]{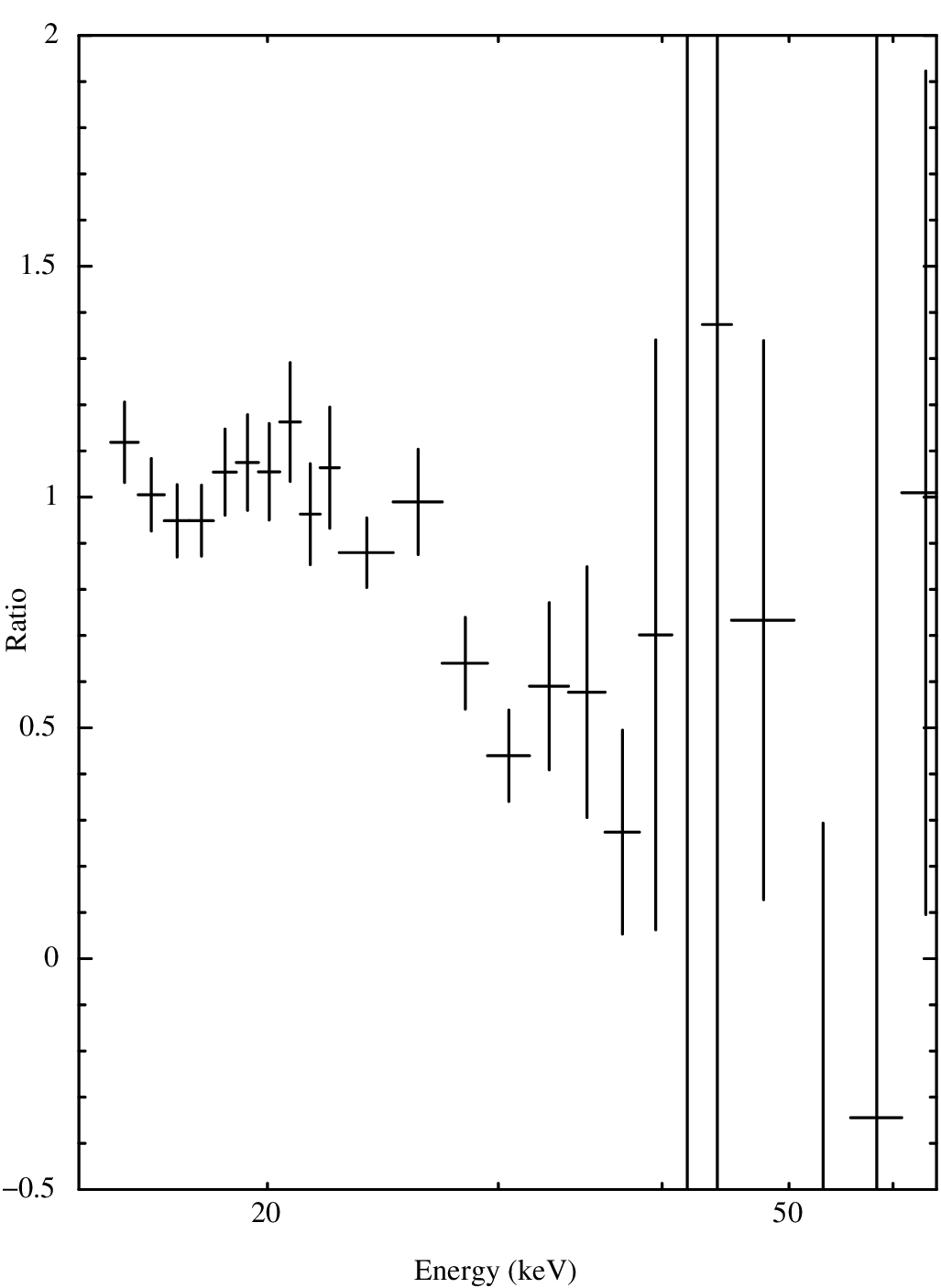}
\caption{Ratio of counts of the energy spectrum with the shallowest cyclotron line to 
counts of the spectrum with the deepest cyclotron line in XTE J1946+274. Though the ratio of the counts after 40 keV have large
error bars due to statistical limitations, the dip in counts at $\sim$ 30-40 keV is clealy visible
indicating the presence of the CRSF}
\label{fig10}
\end{figure}

\begin{figure}
\centering
\includegraphics[scale=0.3,angle=-90]{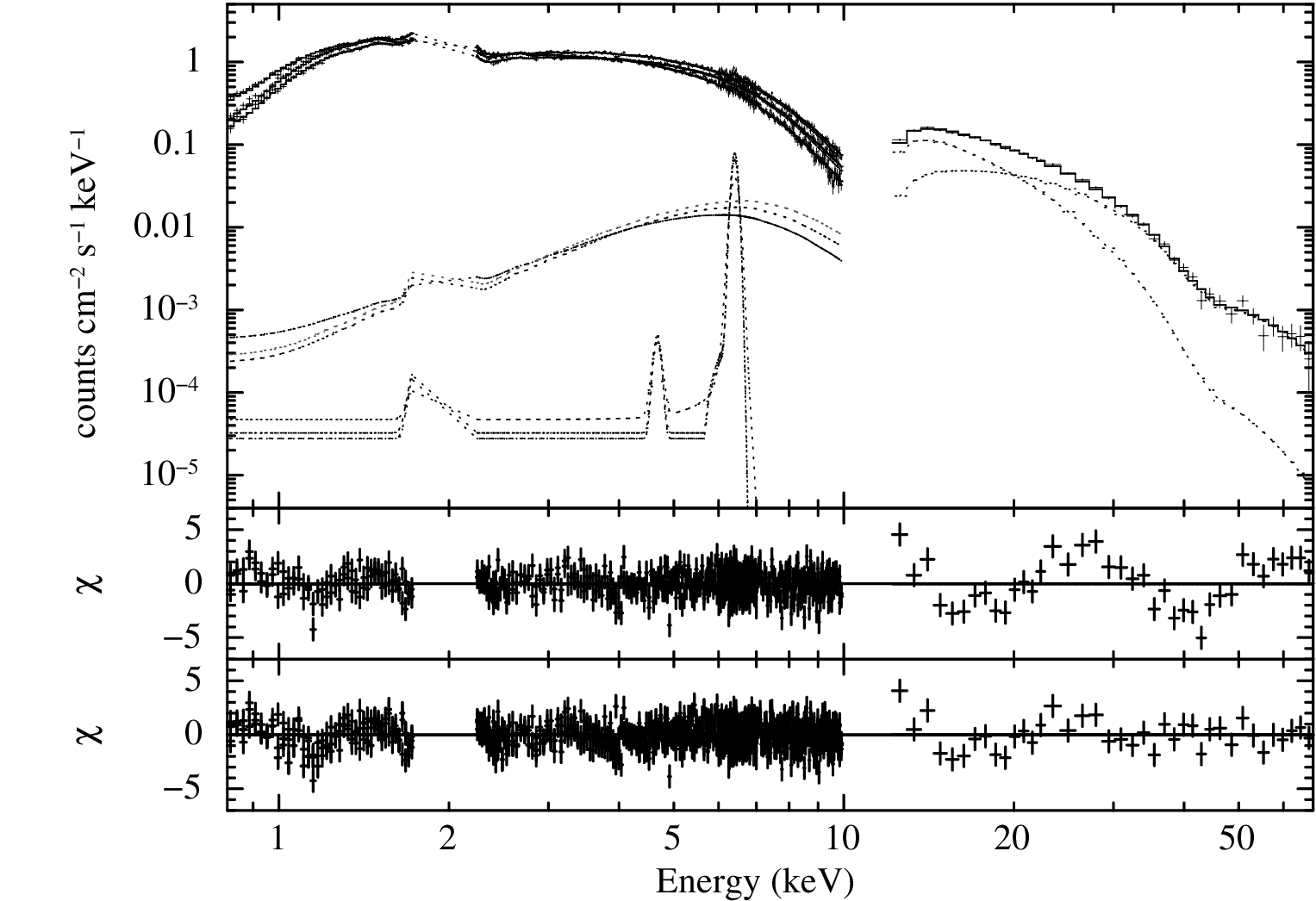} 
\includegraphics[scale=0.3,angle=-90]{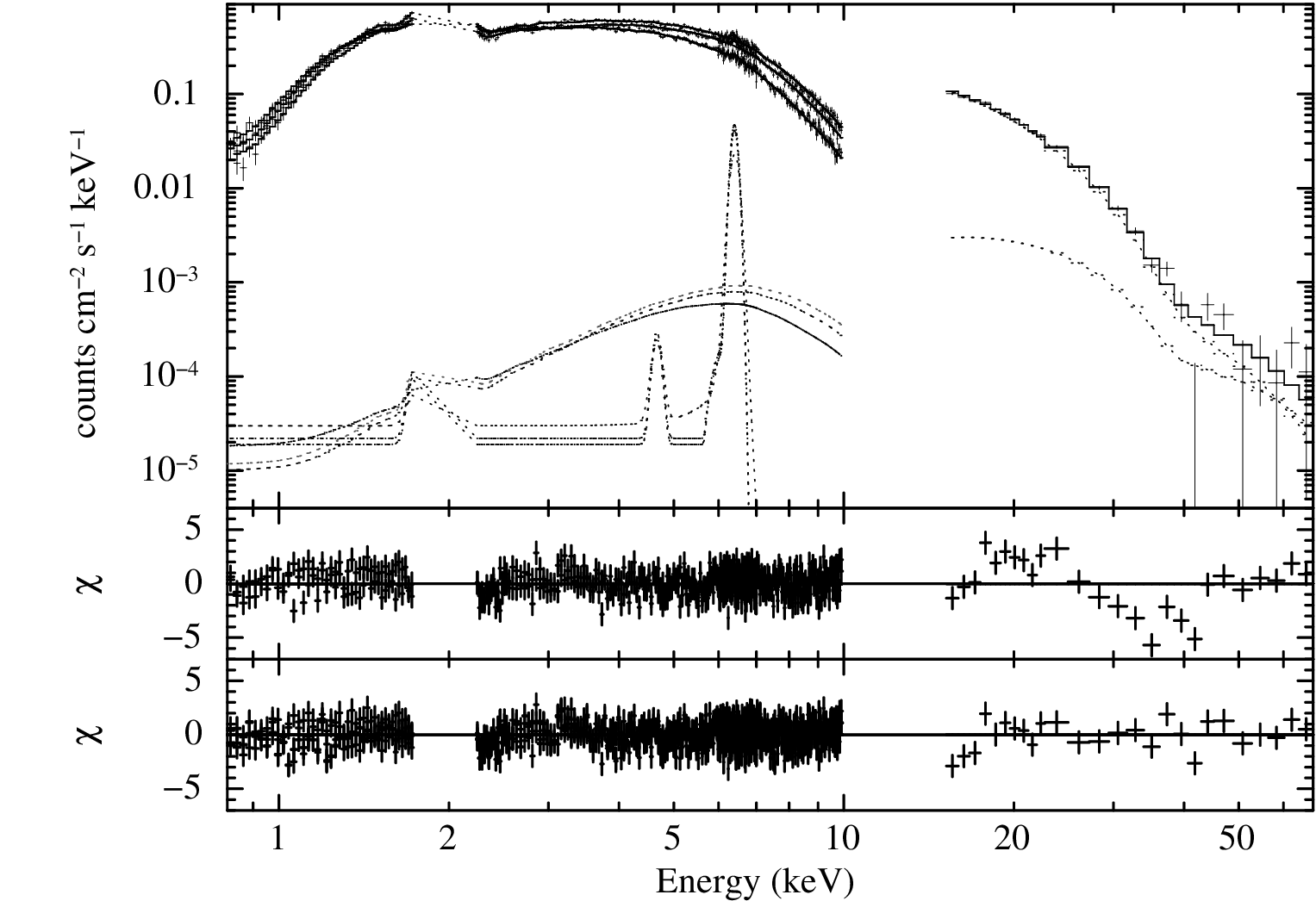} 
\includegraphics[scale=0.3,angle=-90]{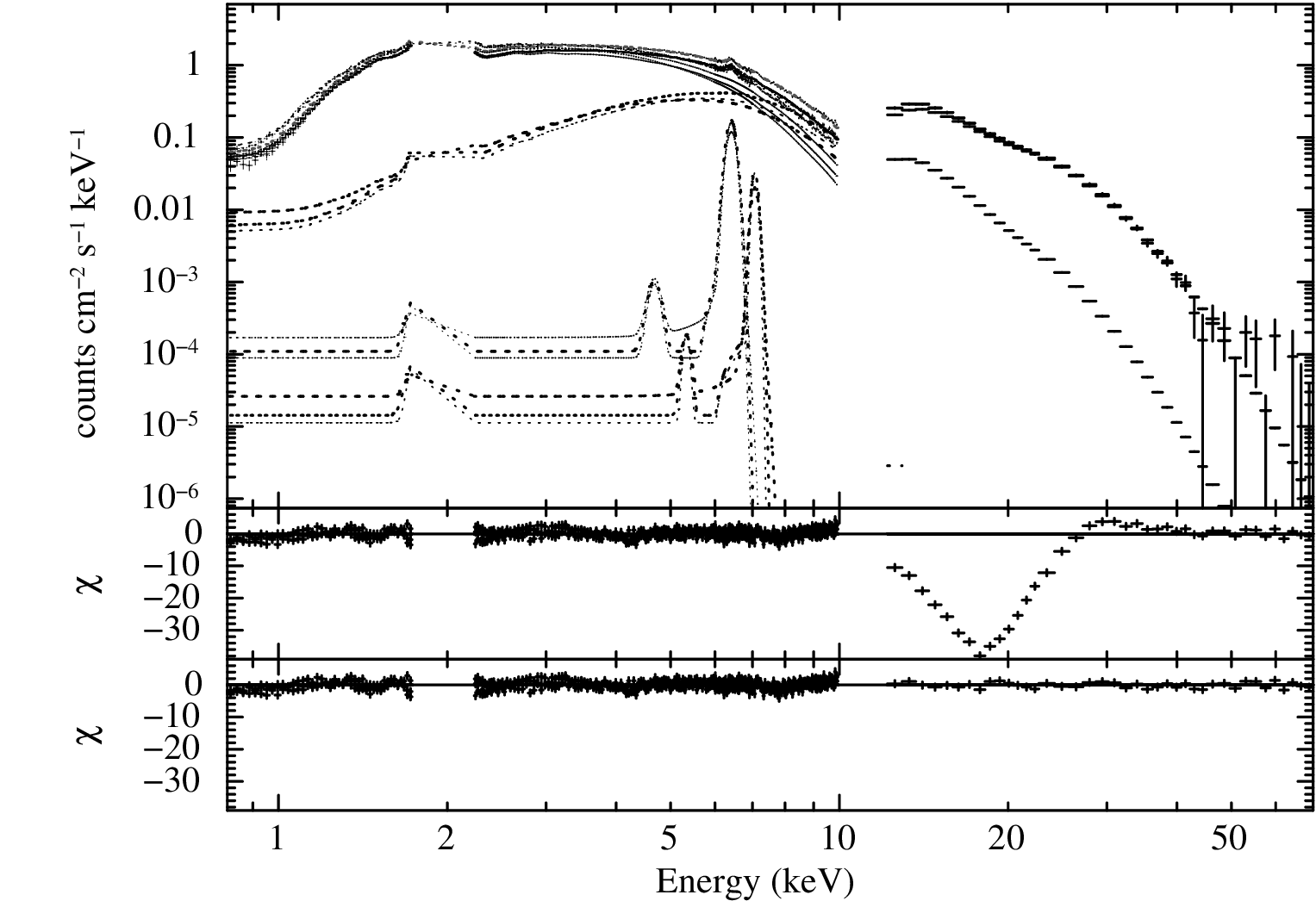} 
\caption{The pulse-phase averaged spectrum of A 0535+26, XTE J1946+274 and 4U 1907+09 showing all
the individual model components (Top left, right and
bottom respectively).
The upper panel shows the best-fit spectra as obtained with the 'NPEX' model. 
The middle panel shows the residuals without inclusion of Lorentzian profile for the CRSF in the spectra,
and the bottom panel shows the residuals after the inclusion of the CRSF. This clearly shows the presence 
of the CRSFs in the energy spectra of all the sources}
\label{fig5}
\end{figure}

\begin{figure}
\centering$
\begin{array}{cc}
\includegraphics[height=12cm, width=6.5cm]{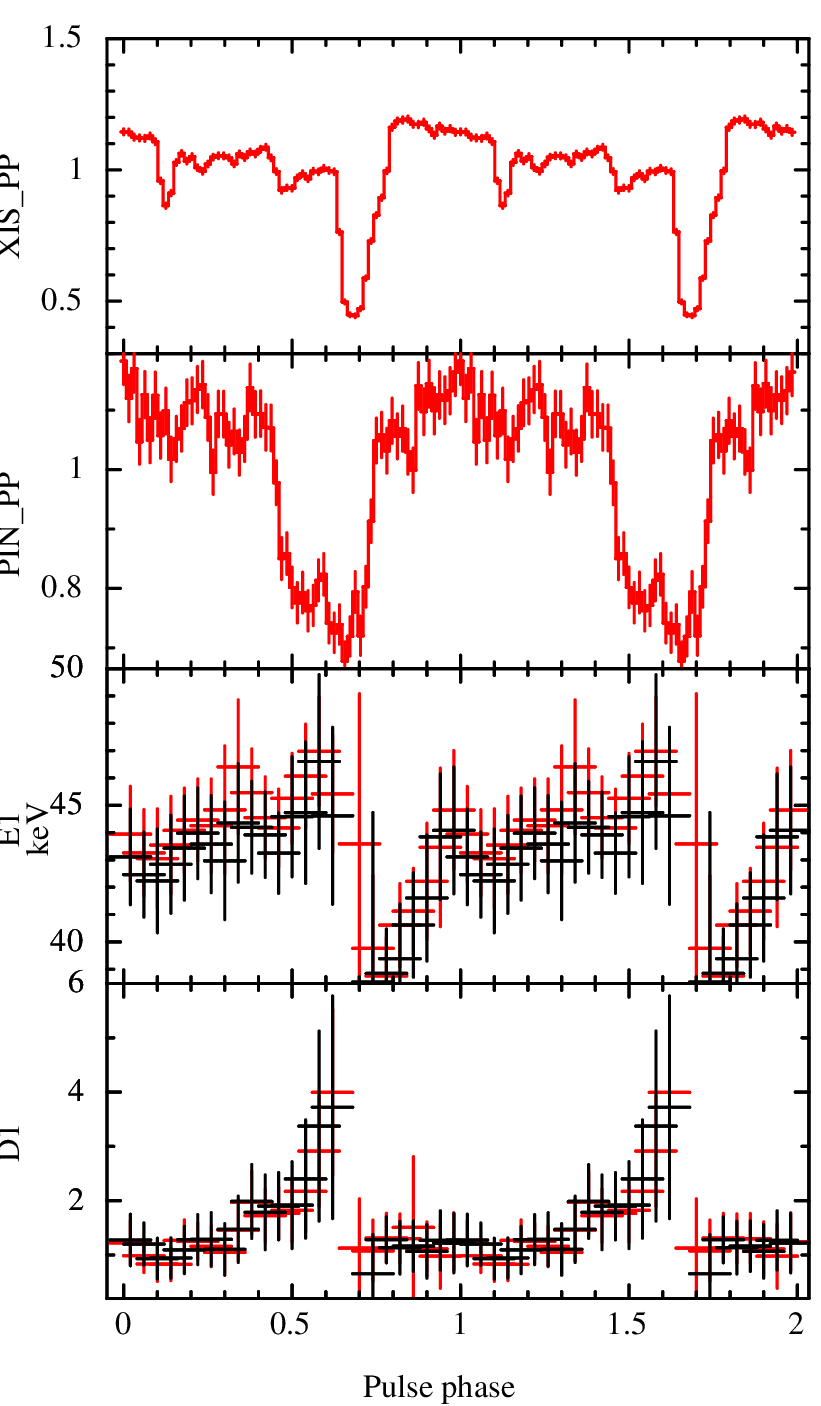} &
\includegraphics[height=12cm, width=8cm]{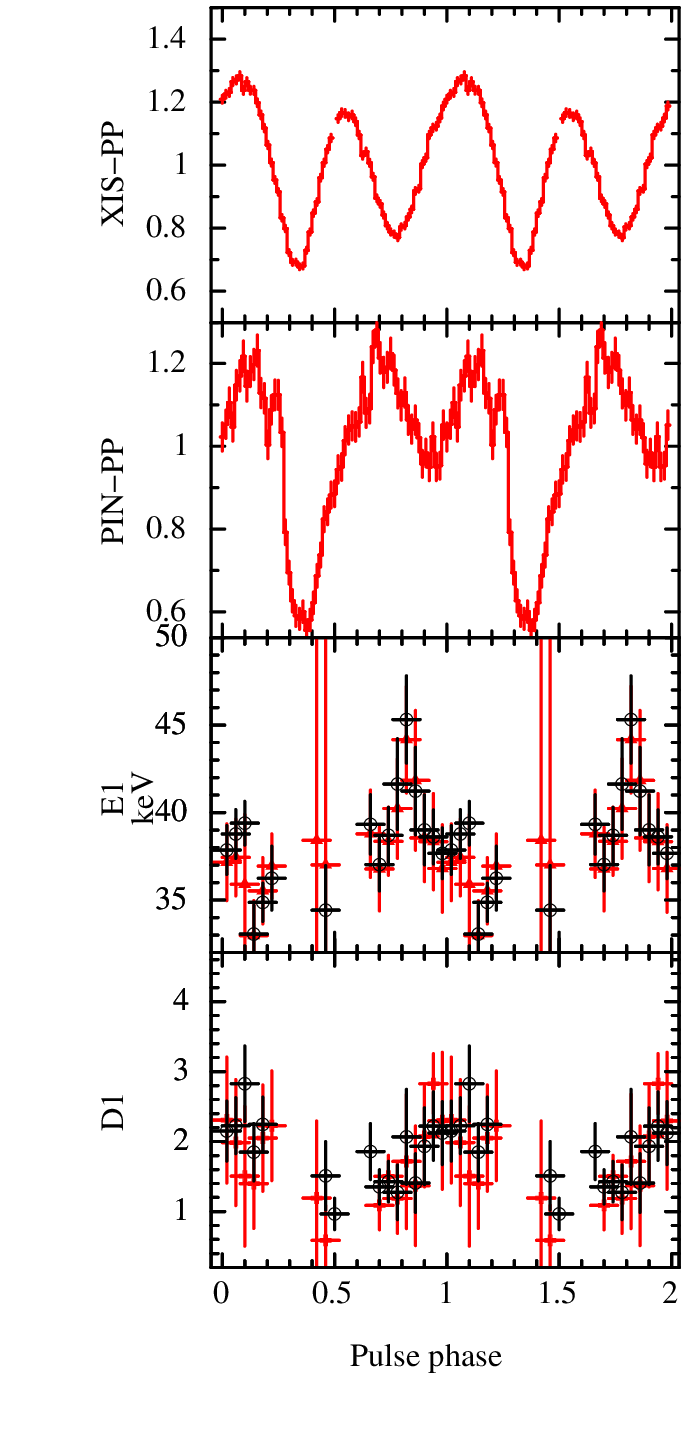} \\
\end{array}$
\caption{Variation of the cyclotron line parameters in A 0535+26 (left panel) and XTE J1946+274 (right panel) as is obtained with the two models. In the left panel, the black points 
 denotes the parameters as obtained with the 'NPEX' model The red points denote the
parameters as obtained with the 'CompTT' model. In the right panel the black points are obtained with the 'highecut' model and the red points with the 'NPEX'
model. Only 8 of the 25 bins are independent. The XIS (0.3-10 keV) and PIN (10-70 keV) pulse profiles are shown in the top
two panels respectively which denote the normalized 
intensity.}
\label{fig6}
\end{figure}

4U 1907+09 is a variable X-ray source showing flaring and dipping activity in the timescales of minutes to hours
\citep{intzand1997}. We performed timing analysis after applying barycentric corrections to the 
event data files using the FTOOLS task 'aebarycen'. Light curves were extracted with a time resolution of 8 s 
(full window mode of the XIS data) for the XISs (0.2--12 keV), and 1 s 
for the HXD/PIN (10--70 keV) respectively. We applied pulse folding and $\chi^{2}$ maximization 
technique to search for pulsations in the XIS data. The source having an eccentric
orbit with a short orbital period, proper correction of the pulse arrival times are required to accurately determine
the pulse period. However, the orbital ephemeris of this source is not known with high accuracy \citep{intzand1998}.
Thus to account for the orbital motion of the binary, we
included a $\frac{dp}{dt}$ term in the fitting, starting with an initial guess consistant with the parameters of the binary,
and iterating for different values of $\frac{dp}{dt}$ to get the maximum $\chi^{2}$. The best fit period corresponding
to this was $441.113 \pm 0.035$ s MJD 54209.43189 with $\frac{dp}{dt}= 3.1\times 10^{-6}$. This 
value obtained is marginally higher than that found by \cite{rivers2010} ($441.03 \pm 0.03
$). However they have not mentioned, taking into account the orbital correction of the pulse arrival times in their work which
might be a reason for this discrepancy. Figures~\ref{fig1} shows the XIS and PIN light
curves along with the hardness ratio. As can be seen from the figure, the light curves show two flaring features
in between and a dip in the last $\sim$ 10 ks of the observation. These features were also mentioned in \citet{rivers2010},
while performing time resolved spectroscopy of the same \emph{Suzaku} observation, and were probed further by them
to investigate the spectral variability with time. The flares may, however also affect our results of pulse phase resolved spectroscopy. We have thus
compared the pulse profiles and the energy spectra in these stretches individually with that from the 
rest of the observation.
Though the pulse profiles look very similar in all the stretches, the energy spectra is harder with an increased
absorption in the last stretch of the observation containing the dip. 
The main aim of this work being
pulse phase resolved spectroscopy to probe the CRSF parameters, we excluded the  stretch of the observation
coincident with the dip in the light curve for further analysis. The arrows in Figures~\ref{fig1} indicate the length of the observation chosen for this work. Pulse profile for this duration of observation was also created in the 
XIS and PIN energy bands as before for A 0535+26 and XTE J1946+274. Due to the absence of low energy dips in this
source however, the energy dependence of the pulse profiles was not investigated further.                                                                                                                              

\subsection{Pulse phase averaged spectroscopy}
Phase averaged spectroscopy was carried out in the same procedure as in A 0535+26 and XTE J1946+274. Best fits 
were obtained with the 'highecut', 'NPEX' and 'compTT' model with comparable values of reduced $\chi^{2}$ and similar
residual patterns. \cite{rivers2010} also obtained similar results with the 'highecut', 'fdcut' and 'NPEX' model.
A comparison between the ’NPEX’ model parameters obtained in our analysis
and those reported in \cite{rivers2010} reveal a softer less absorbed
spectra obtained by us. This is expected, since we have excluded the
the last stretch of data from our analysis which had a more harder and absorbed
spectra. Two gaussian lines were also used to model the narrow Fe k$\alpha$ and Fe k$\beta$ feature found at 6.4 and
7.1 keV respectively. In addition, a relatively shallow and  narrow feature found at $\sim$ 18 keV was modeled with a
Lorentzian profile which is the CRSF previously detected in this source \citep{makishima1992,makishima1999}. 
As also discussed in \cite{rivers2010}, the first harmonic of the CRSF at $\sim$ 36 keV could not be detected
in the PIN spectra probably due to the statistical limitation of the data in this energy range.
The CRSF parameters obtained with the 'NPEX' and 'CompTT' models were consistant within error bars
with that found by \cite{rivers2010}
who performed phase resolved spectroscopy in 6 independent bins using the gaussian absorption model
(keeping in mind that the centroid
energy of the Lorentzian description is not coincident with the minimum of the line profile \citep{nakajima2010}). The 'highecut' model however required a deeper CRSF to fit the
spectra. The reduced $\chi^{2}$ obtained for the models
 were 1.62, 1.51 and 1.69 for 832, 837 and 838 d.o.f for highecut, NPEX and CompTT respectively with no systematic 
 residual patterns.
Due to the compatibility of the CRSF parameters obtained with the 'NPEX' and CompTT models, we have carried out further phase
resolved analysis using these two models. Figure \ref{fig5} shows the best-fit spectra
for 4U 1907+09 along with the
 residuals before and after including
 the CRSF, thus showing the presence of the feature clearly. The CRSF in this source is very strong and has also
 been reported in the same \emph{Suzaku} observation before \citep{rivers2010}. We therefore do not comment
 on its detection significance.

 \subsection{Pulse phase resolved spectroscopy}
For investigating the pulse phase-resolved spectroscopy of the CRSF, we generated phase resolved spectra with
25 overlapping but 8 independent phase bins and used the same analysis procedure as discussed previously for
A 0535+26 and XTE J1946+274.
We were however able to constrain the phase dependent variation of all the CRSF parameters for this source, probably
due to the longest observation duration available for it and a low cyclotron energy compared
to the other sources. Figure \ref{fig6} shows the variation of the cyclotron parameters of the source using the 
best fit models as a function of pulse phase. As in the case of the previous sources, 
the similar pattern of variation obtained for the different continuum models used give us considerable 
amount of confidence on the obtained results. The following characteristics can be observed 
in more detail from Figure \ref{fig6}. As before, the variations are compared with respect to the high energy PIN profile.
\begin{enumerate}
 \item The energy \ensuremath{E1_{\mathrm{cycl}}} varies by $\sim 19\%$. Its value is maximum near the peak of the first pulse
 (20 keV at phase $\sim$ 0.3), and again at the ascending edge of the second pulse (phase $\sim 0.6$),
 the minimum being at the second pulse peak (15 keV at $\sim$ phase 0.7--0.8).
 \item The depth \ensuremath{E1_{\mathrm{cycl}}} has a clear double peaked pattern with the peaks corresponding to the ascending edge
 of the first pulse and the peak of the second pulse (phase $\sim 0.1-0.2$ and 0.7 respectively). It is
 minimum near the pulse minima (phase $\sim$ 0.9). \ensuremath{E1_{\mathrm{cycl}}} varies between 0.2--1.4 and is generally greater for the first pulse. 
 \item The width (\ensuremath{W1_{\mathrm{cycl}}}) has a similar pattern of variation as \ensuremath{E1_{\mathrm{cycl}}}, and peaks at similar phases with values varying
 within 7 keV.  
\end{enumerate}
 
\section{Discussions \& Conclusions}
In the present work we have presented the results of detailed pulse phase resolved spectroscopy of the CRSF
parameters of A 0535+26, XTE J1946+274 and 4U 1907+09 using long \emph{Suzaku} observations.
Pulse phase dependence of the CRSF parameters are obtained for the first time in A0535+26
and XTE J1946+274 and a more detailed and careful analysis has been done in 4U 1907+09
which is consistant with the earlier result obtained using the same observation \citep{rivers2010}.
The analysis is done taking into account various factors which might smear the pulse phase dependence 
results as mentioned in earlier sections. The strength of our results lie in the fact that we
have obtained similar pattern of variation of the CRSF parameters for all the three sources with more than
one continuum model.
\subsection{pulse phase dependence of the cyclotron parameters}
Results of pulse phase resolved spectroscopy of the cyclotron parameters have been presented previously in
some sources, 
for example in Her X-1 \citep{soong1990,enoto2008,klochkov2008}, 4U 0115+63 
\citep{heindl2000}, Vela X-1 \citep{kreykenbohm1999,kreykenbohm2002,labarbera2003,maitra2012}, 4U 1538-52 
\citep{robba2001}, Cen X-3 \citep{suchy2008}, 
and more recently in GX 301-2 \citep{suchy2012}, 1A 1118-61 \citep{suchy2011,maitra2011} 
and 4U 1626-67 \citep{iwakiri2012}. \\
By modeling the pulse phase dependence with different continuum models, we have been able to establish the
robustness of the results. In the process of trying to fit the energy spectrum with different continuum models
we have also noticed certain trends in continuum model fitting. The 'highecut' model being a very simple
model with less number of parameters, is a good choice to model the continuum in case of moderate or poor
statistics. This is evident in the case of XTE J1946+274. The 'CompTT' on the hand which is a 
more physical description of the spectra and has a reasonable number of free parameters is better for
continuum fitting specially for phase resolved spectroscopy if the statistical quality of the data is reasonably
good. This is probably the reason why it failed to constrain the continuum well in the case of XTE J1946+274.
The 'NPEX' model approximates the photon number spectrum for an unsaturated Comptonization
\citep{sunyaev1980,meszaros1992}, and has a clear physical meaning inspite of being a phenomenological model. It
is useful for all the three sources with significantly different
statistical quality.

 By assuming certain physics and geometry of the line forming region, the CRSF feature has been modeled 
 analytically and with simulations by \citet{araya1999,araya2000,schonherr2007} and more recently by 
\citet{nishimura2008,nishimura2011,mukherjee2011}. Although these models predict  variations in the depth, 
width and the centroid
energy of the CRSF features with the changing viewing angle at different pulse phases, a variation
in the CRSF parameters as large as $30 \%$ as found in our results needs to take into account either
a possible deviation or distortion from the simple dipole geometry of the magnetic field \citep{schonherr2007,mukherjee2011}
, a gradient in the
field itself \citep{nishimura2008}, or a different geometry of the accretion column \citep{kraus2001}. A detailed
modeling taking into account these factors would provide us a detailed information about the geometry and
emission patterns of the sources. However
simpler interpretations can be done, since the correlation of the deepest and shallowest CRSFs with the pulse profile of the source
can provide some idea about the beaming pattern of the source at that luminosity.
Following this, the trend of shallowest lines near the pulse peak and deepest near the off-pulse as found in A 0535+26
and XTE J1946+274, favors a pencil beam geometry. On the other hand, deepest and widest lines found near the peak
and shallowest and narrowest near the off-pulse as found in 4U 1907+09 favors a fan beam geometry for the emission.
These results
may be further complicated by assuming the contribution from both the magnetic poles of the neutron star in contrary 
to one of them, either due to gravitational light bending
or particular geometry of the system allowing the view of both the poles. Modeling of the variations of the CRSF parameters with pulse phase is ongoing. Detailed
discussions on the same will be made in a future work.
 
\acknowledgments

This research has made use of data obtained through the High Energy Astrophysics Science Archive Research Center On line Service,
 provided by NASA/Goddard Space Flight Center. Chandreyee Maitra would like to thank Carlo Ferrigno for providing the 'fdcut'
  and the 'newhighecut' local models.


\begin{figure}
\centering
\includegraphics[height=12cm, width=8cm]{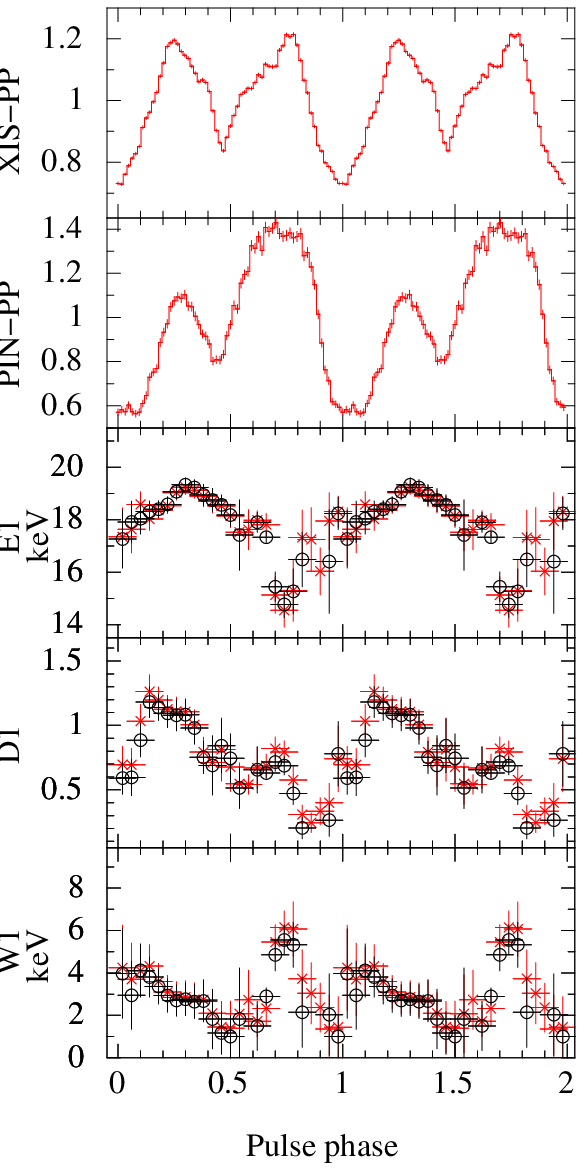}
\caption{Variation of the cyclotron line parameters in 4U 1907+09 as is obtained with the two models. The black points 
 denotes the parameters as obtained with the 'NPEX' model The red points denote the
parameters as obtained with the 'CompTT' model. Only 8 of the 25 bins are independent. The XIS (0.3-10 keV) and PIN (10-70 keV)pulse profiles are shown in the top
two panels respectively which denote the normalized 
intensity.}
\label{fig9}
\end{figure}

\begin{figure}
\centering$
\begin{array}{cc}
\includegraphics[height=10cm, width=6cm]{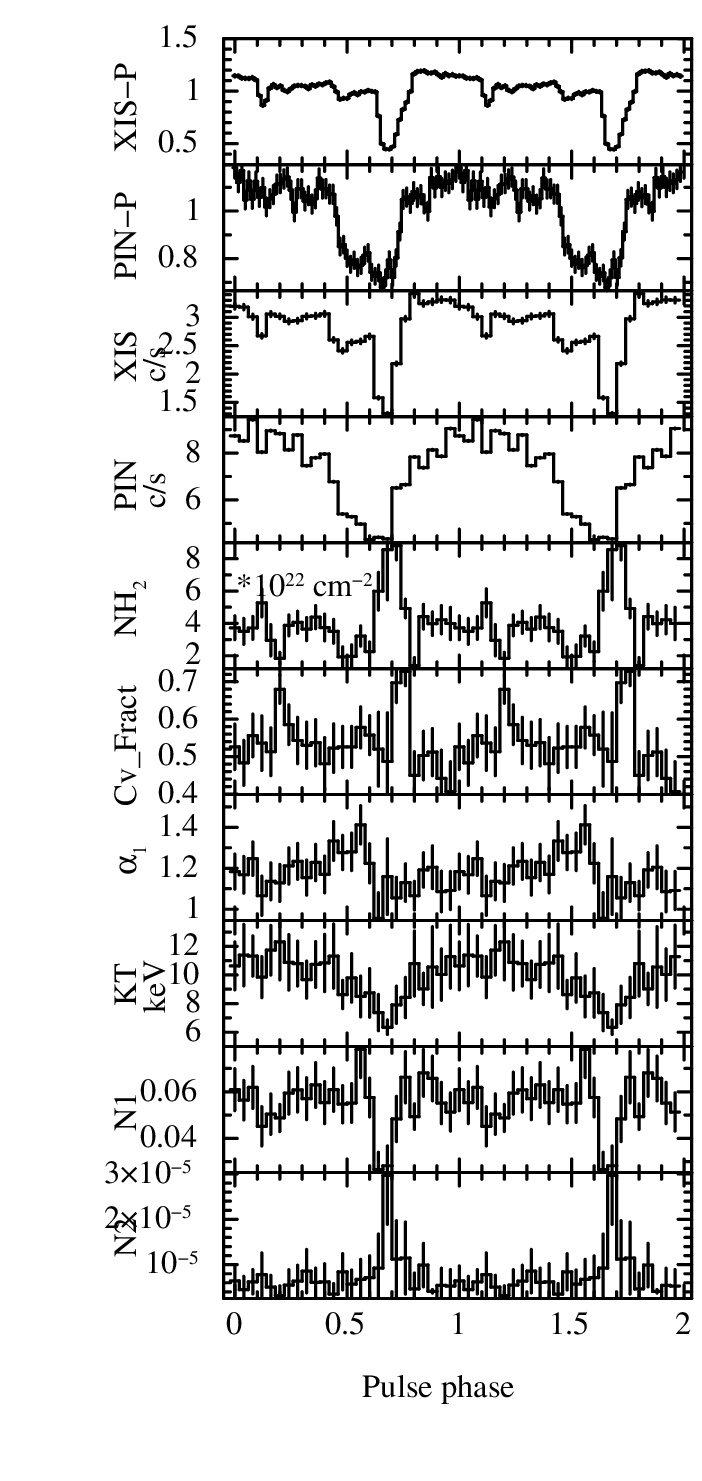} &
\includegraphics[height=10cm, width=7cm]{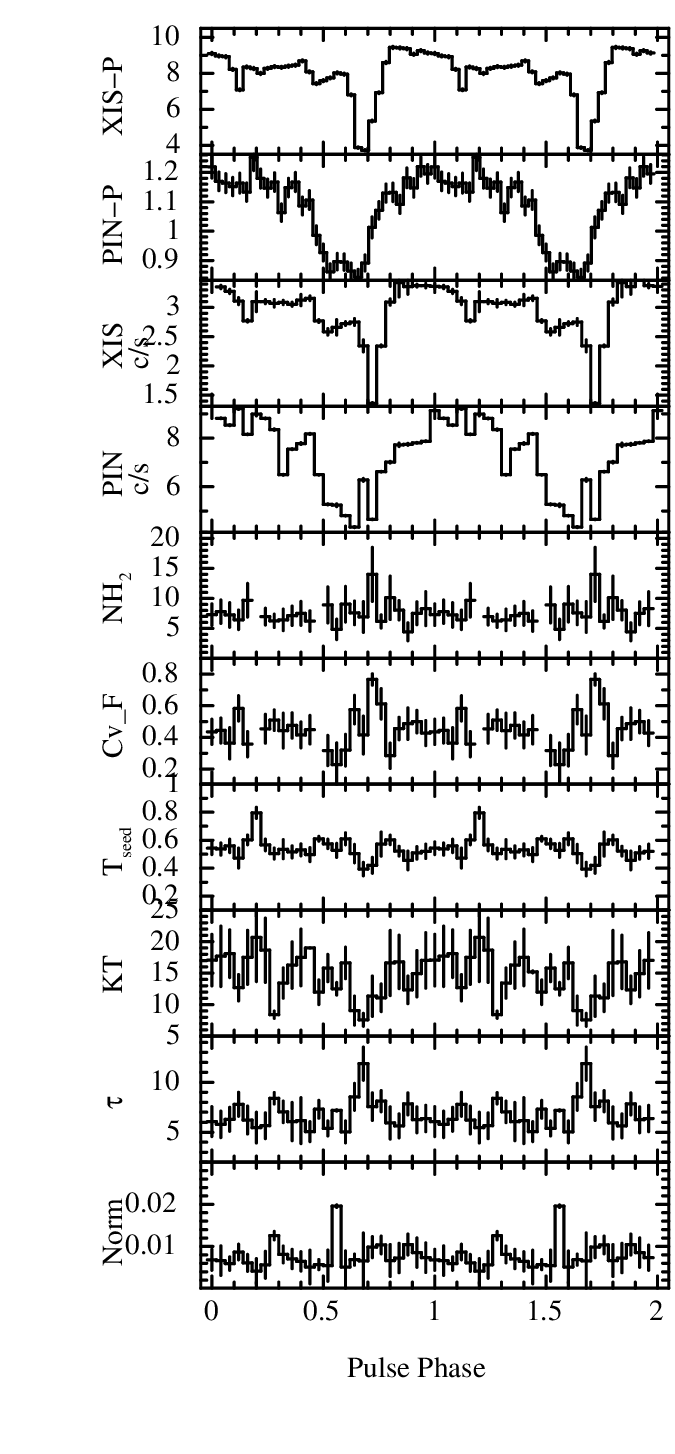} \\
\end{array}$
\caption{Variation of the spectral parameters with phase along with the pulse profile (0.3-12
keV for XIS and 10-70 keV for PIN) in A 0535+26. The left panel shows the variation using the model
'NPEX' and the right panel shows the same using the model 'CompTT'. The pulse profiles denote the normalized 
intensity. The XIS (0.3-10 keV) and PIN (10-70 keV) pulse profiles are shown in the top
two panels respectively which denote the normalized 
intensity.}
\label{fig7}
\end{figure}

\begin{figure}
\centering$
\begin{array}{cc}
\includegraphics[height=10cm, width=6cm]{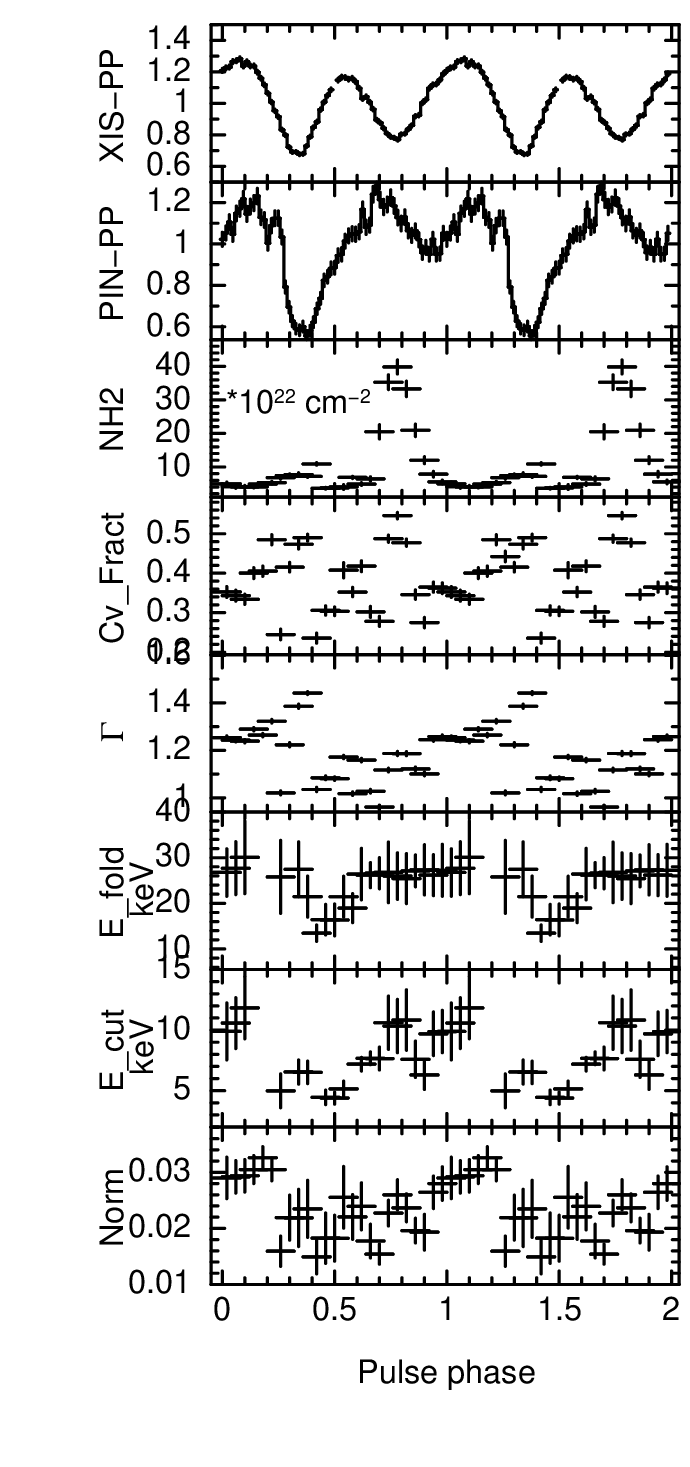} &
\includegraphics[height=10cm, width=6cm]{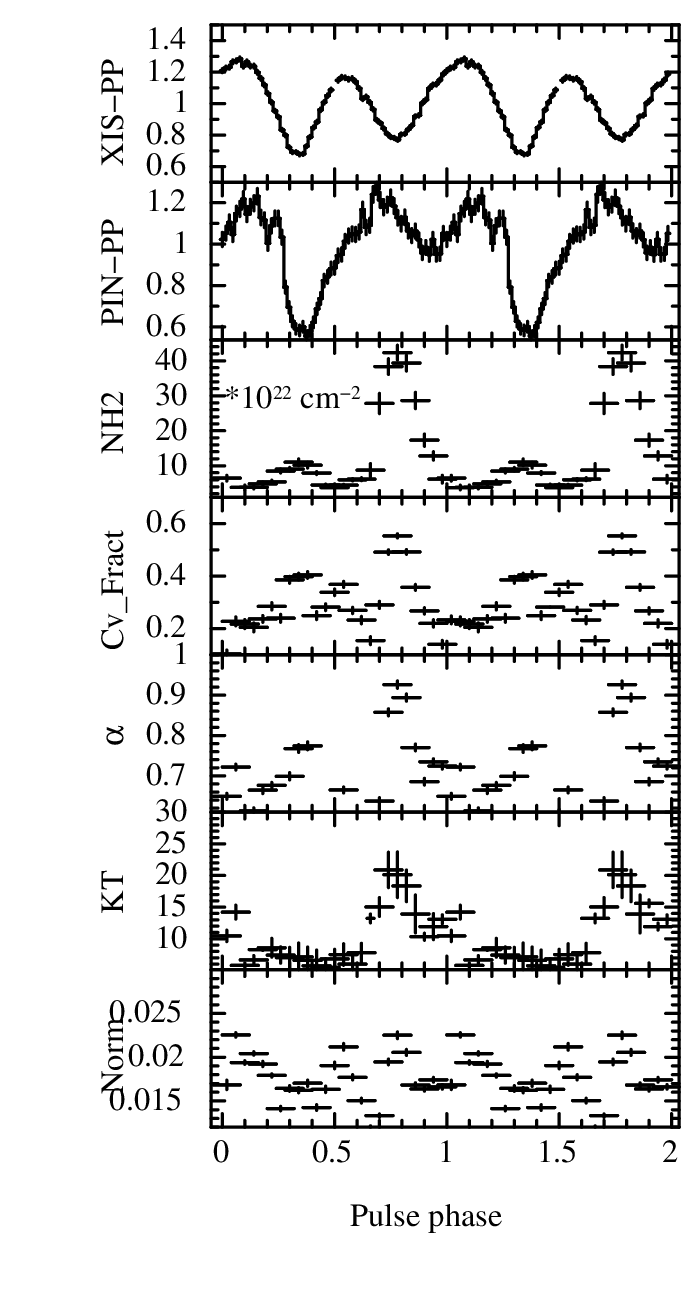} \\
\end{array}$
\caption{Variation of the spectral parameters with phase along with the pulse profile (0.3-12
keV for XIS and 10-70 keV for PIN) in XTE J1946+274. The left panel shows the variation using the model
'highecut' and the right panel shows the same using the model 'NPEX'.The pulse profiles denote the normalized 
intensity. The XIS (0.3-10 keV) and PIN (10-70 keV) pulse profiles are shown in the top
two panels respectively which denote the normalized intensity.}
\label{fig8}
\end{figure}

\begin{table*}[p]
\caption{Best fitting phase averaged spectral parameters of A 0535+26, XTE J1946+274 and 4U 1907+09. Errors quoted are for 99 per cent confidence range.}
\scriptsize 
\begin{tabular}{c c c c}
\hline \hline
Model &  A 0535+26 & XTE J1946+274 & 4U 1907+09 \\
parameters & NPEX \quad CompTT & highecut \quad NPEX & \quad NPEX \quad CompTT \\
\hline
$\ensuremath{N_{\mathrm{H1}}}^{a}$ ($10^{22}$ atoms $cm^{-2}$) & $0.59 \pm 0.02$  \quad $0.31_{-0.03}^{+0.04}$ & $1.30 \pm 0.028 $ \quad $1.27_{-0.04}^{+0.03} $& $1.97 \pm 0.01$ \quad $1.62 \pm 0.03 $\\
$\ensuremath{N_{\mathrm{H2}}}^{b}$ ($10^{22}$ atoms $cm^{-2}$)  & $3.74_{-0.36}^{+0.25}$ \quad $7.41_{-1.13}^{+1.45}$ & $5.86_{-0.82}^{+0.87}$ \quad$9.03_{-1.73}^{+2.72}$ & --\quad --\\
\ensuremath{Cv_{\mathrm{fract}}} & $0.49 \pm 0.01 $ \quad $0.37 \pm 0.02$  & $0.27 \pm 0.04  $ \quad $ 0.17_{-0.05}^{+0.04} $ & --\quad --\\
PowIndex & -- \quad --& $1.09_{-0.04}^{+0.05}$\quad -- & -- \quad --\\
E-folding energy (keV) & -- \quad --& $25.57_{-2.26}^{+2.75} $ \quad-- & --\quad --\\
E-cut energy (keV) & --\quad -- & $7.02_{-0.29}^{+0.69} $ \quad-- & -- \quad --\\
$\ensuremath{powerlaw_{\mathrm{norm}}}^{c}$ & -- \quad -- & $0.021 \pm 0.001 $ \quad -- & --\quad --\\
\ensuremath{CompTT_{\mathrm{T0}}} (keV) & --\quad $0.52 \pm 0.04$& --\quad --& --\quad $0.47 \pm 0.01$\\
CompTT KT (keV) & --\quad $ 14.67_{-1.53}^{+2.07}$& --\quad --& --\quad $ 4.78 \pm 0.04$ \\
CompTT $\tau$ & --\quad $7.33_{-0.48}^{+0.44}$ & --\quad --&--\quad $14.87 \pm 0.08 $\\
\ensuremath{CompTT_{\mathrm{norm}}} $^{c}$ & -- \quad $0.008 \pm 0.0009$&--\quad --& --\quad $0.0317 \pm 0.0005 $ \\
NPEX $\alpha1$  & $1.01_{-0.04}^{+0.05}$ \quad --& --\quad $ 0.70_{-0.10}^{+0.08}$ & $0.84 \pm 0.02 $\quad --           \\
NPEX $\alpha2$  & -2.0 (frozen) \quad -- & --\quad -2.0 (frozen) & -2.0 (frozen) \quad -- \\
NPEX KT (keV)  & $11.49_{-1.13}^{+1.56}$ \quad -- & --\quad $13.29_{-2.74}^{+2.43}$ & $4.41 \pm 0.04$ \quad --           \\
 \ensuremath{NPEX_{\mathrm{norm}}} $1^{c}$  & $0.041_{-0.001}^{+0.002}$ \quad -- & --\quad $0.014_{-0.001}^{+0.002}$ & $0.072 \pm 0.001$ \quad --\\
 \ensuremath{NPEX_{\mathrm{norm}}} $2^{c}$  & $4.94e-6 \pm 1.25e-6$  \quad -- & -- \quad $2.34^{d}$ $\pm$ $1.5$  &  $2.47e-4$ $\pm$ $7.5e-6$  \quad -- \\
\ensuremath{E1_{\mathrm{cycl}}} & $42.60_{-0.84}^{+0.91}$ \quad $43.24_{-0.65}^{+0.85} $ & $38.30_{-1.36}^{+1.63} $\quad $38.65_{-1.76}^{+1.97}$& $17.96_{-0.19}^{+0.20}$ \quad $18.07 \pm 0.18$ \\
 \ensuremath{D1_{\mathrm{cycl}}} $^{e}$ & $1.37_{-0.23}^{+0.25}$ \quad $1.43_{-0.26}^{+0.35}$ & $1.72_{-0.28}^{+0.41}$ \quad $1.50_{-0.43}^{+0.56}$ & $ 0.68 \pm 0.03$ \quad $0.61 \pm 0.03 $\\
\ensuremath{W1_{\mathrm{cycl}}} & $7.17_{-1.82}^{+2.28}$ \quad $4.93_{-1.20}^{+1.37}$ & $ 9.61_{-3.06}^{+3.69}$\quad $8.61_{-3.32}^{+5.45}$ & $3.34_{-0.37}^{+0.39}$\quad $2.78_{-0.37}^{+0.39}$\\
Iron line energy (keV) & --\quad --& --\quad--& $7.10_{-0.024}^{+0.035}$\quad $7.10_{-0.037}^{+0.033} $\\
Iron line eqwidth (eV) & --\quad-- & --\quad --& $ 10.74 \pm 2.90$ \quad $ 7.9 \pm 2.6$\\
Iron line energy (keV) & 6.41$\pm$ 0.01 \quad 6.41$\pm$ 0.01& $6.41 \pm 0.022$ \quad $6.41 \pm 0.02$& $6.42_{-0.010}^{+0.008}$ \quad $6.42_{-0.009}^{+0.007} $ \\
Iron line eqwidth (eV) & $23.37 \pm4.61 $ \quad $24.28 \pm5.26$ &  $29.05 \pm 4.93$ \quad $28.39 \pm 5.15 $  & $51.67 \pm 3.62 $ \quad $43.81 \pm 1.66 $\\
Flux (XIS) $^f$ (0.3-10 keV) &  $3.05 \pm 0.04$\quad $ 3.05 \pm 0.05$ & $1.80 \pm 0.02$ \quad $1.80 \pm 0.02$& $4.75 \pm 0.06 $ \quad $4.74 \pm 0.06 $\\
Flux (PIN) $^g$ (10-70 keV) &  $8.14 \pm 0.02$\quad $8.14 \pm 0.03$& $3.47 \pm 0.01 $\quad $3.59 \pm 0.0 $ & $6.87 \pm 0.02 $ \quad $6.91 \pm 0.02$\\
reduced $\chi^{2}$/d.o.f & 1.25/839\quad1.26/837 & 1.09/826\quad1.11/826& 1.51/837\quad 1.69/838\\
\hline
\end{tabular}\\
$^a$ Denotes the Galactic line of sight absorption
$^b$ Denotes the local absorption by the partial covering absorber 'pcfabs'.\\
$^c$ \ensuremath{\mathrm{photons}\, \mathrm{keV}^{-1}\,\mathrm{cm}^{-2}\,\mathrm{s}^{-1}\,\mathrm{at}\, 1\, \mathrm{keV}}\\
$^d$ $*10^{-7}$ ; $^c$ denotes the optical depth of the feature. \\
$^e$ \ensuremath{10^{-10}\,  \mathrm{ergs}\,  \mathrm{cm}^{-2}\,\mathrm{s}^{-1}} and are in 99 \% confidence range.\\
$^f$ \ensuremath{10^{-10}\,  \mathrm{ergs}\,  \mathrm{cm}^{-2}\,\mathrm{s}^{-1}} and are in 99 \% confidence range.
\end{table*}

\end{document}